\documentclass[article]{jfm}
\usepackage{graphicx}
\usepackage{newtxtext}
\usepackage{newtxmath}
\usepackage{natbib}
\usepackage{hyperref}
\hypersetup{
    colorlinks = true,
    urlcolor   = blue,
    citecolor  = black,
}

\newcommand{\RomanNumeralCaps}[1]
\linenumbers

\usepackage{amsmath}
\usepackage{bm}
\usepackage{subfigure}
\usepackage{xcolor}
\usepackage{mathtools}
\usepackage{xr-hyper} 
\usepackage{upgreek}
\usepackage{cooltooltips}
\usepackage[normalem]{ulem}

\newcommand{\modDL}[1]{}
\newcommand{\modTY}[1]{}

\title{Stabilising nonlinear travelling waves in pipe flow using time-delayed feedback}
\author{Tatsuya Yasuda\aff{1}\corresp{\email{ty44@st-andrews.ac.uk}} 
\and Dan Lucas\aff{1}\corresp{\email{dl21@st-andrews.ac.uk}}}
\affiliation{\aff{1}
School of Mathematics and Statistics, 
University of St Andrews, Mathematical Institute, North Haugh, 
St Andrews KY16 9SS, UK}

\begin{document}
\maketitle

\begin{abstract}
We demonstrate the first successful non-invasive stabilisation of nonlinear travelling waves in a straight \modDL{cylindrical} pipe 
using time-delayed feedback control (TDF) \modDL{working in various symmetry subspaces}. 
\end{abstract}




\section{Introduction}\label{sec:introduction}
%
In this study, we consider the incompressible Navier--Stokes equations as a high-dimensional dynamical system, where 
simple invariant solutions, 
or exact coherent structures (ECSs), can be considered key building blocks of the spatiotemporal chaos \citep[see, e.g.,][]{Kawahara:2012iu,graham_2021_exact}. 
Exact coherent structures take the form of equilibria, 
relative equilibria (or travelling waves), periodic orbits, relative periodic orbits 
and \modDL{ invariant torii}. 
The study of ECS has shed light on the origin of turbulence statistics 
\citep{Chandler,Lucas:2015gt,Page_2024}, 
 the physical mechanisms that play a role in sustaining turbulence 
\citep{vanVeen:2006fm,Lucas:2017fz,yasuda_2019_a,graham_2021_exact, McCormack_Cavalieri_Hwang_2024}, subcritical transition to turbulence \citep{Skufca_2006,Kreilos:2012bd}, mixing and layer formation in stratified flow \citep{LC,LCK}, pattern formation in convection \citep{Beaume:2011kh,Reetz_Schneider_2020} and drag reduction \citep{Bengana_Yang_Tu_Hwang_2022} amongst other applications. 
The fundamental objective is to be able to describe a turbulent flow by using ECSs as a reduced order description and as a simple way to predict its statistics. 
ECSs in various fluid flows have been successfully isolated using homotopy \citep{Nagata:1990ei}, 
bisection \citep{2001JPSJ...70..703I,duguet_2008_transition} and \modDL{quasi-Newton shooting} \citep{Viswanath:2007wc,Viswanath:2009vu,Chandler}. 
Newton-Krylov \modDL{``shooting'' methods} have proved the most successful and efficient methods for converging these unstable states \modDL{from the turbulent attractor}, however, as the Reynolds number or system size are increased, and the flow becomes increasingly disordered, well-conditioned guesses become harder to identify. \modDL{Recent research has sought to address this issue, generally by identifying better starting guesses (e.g. using dynamic mode decomposition \citep{page_kerswell_2020,Marensi_Yalniz_Hof_Budanur_2023},  convolutional autoencoders \citep{page2023exact} or improved recurrence conditions \citep{Redfern_2024}) and/or by `preprocessing' guesses by gradient descent (e.g. using automatic differentation \citep{Page_2024} to obtain derivatives).  An alternative to shooting is to converge closed loops using variational methods to minimise cost functions \citep{azimi2020,Parker_Schneider_2022} which ensure the governing equations are satisfied. This approach has larger radius of convergence but is much slower to converge than Newton shooting.}
%
%

Here, \modDL{our approach will be to dispense with an iterative or root-finding approach} and instead control the underlying ECSs thereby \modDL{allowing one to simply timestep a modified set of equations} onto them. We will develop a time-delayed feedback control method to stabilise travelling waves from turbulence in pipe flow. This method serves to complement existing methods to find ECSs and provides new insight into flow control in general. 
%

A relatively rich literature on ECSs in pipe flow exists, with many travelling wave solutions
\citep{faisst_2003_traveling, 
wedin_2004_exact,
pringle_2009_highly,
Viswanath:2009vu,
Willis:2013bu, 
ozcakir_2016_travelling,
ozcakir_2019_nonlinear} and
relative periodic solutions
\citep{budanur_2017_relative,duguet_2008_relative} reported. 
%
%
Moreover experimental observations suggest that these solutions to the governing equations, \modDL{even} with idealised boundary conditions \modDL{(periodicity in the streamwise direction),} do have relevance to real-world applications \citep{Hof:2006fk}. 
Feedback control has been used to great effect by \citet{willis_2017} to obtain `edge-states' in pipe flow. \modDL{However, when applied to channel flow, \citet{linkmann_2020} discovered that such control can induce new instabilities, even when the original unstable mode is stabilised.} 
It remains to be seen if more generalised control methods can obtain other ECSs by stabilising more than one direction. For these reasons pipe flow is an excellent candidate system in which to test and develop this control approach. 

%
%
%

%
The method of time-delayed feedback \modTY{(TDF)} control, sometimes called Pyragas control \citep{Pyragas:1992ch}, is a well-known approach for stabilising invariant solutions in chaotic systems and has been used to great effect in a variety of dynamical systems \citep{Ushakov:2004em,Popovych:2005hj,Yamasue:2006,Stich:2013,Luthje:2001do} .
%
A finite-dimensional \modDL{autonomous} dynamical system with state vector $\bm{X}(t)$ can be expressed as 
\begin{eqnarray}\label{eq:pyragas1}
\frac{\mathrm{d}\bm{X}}{\mathrm{d}t} = \bm{f}(\bm{X}(t); \bm{p}) + \bm{F} \:,
\end{eqnarray}
in our case the vector field $\bm{f}$ is given by  the discretised Navier-Stokes equation with parameters $\bm{p}$, and we denote $\bm{F}$ as the time-delayed feedback `force'; 
\begin{eqnarray}\label{eq:pyragas2}
\bm{F}(t) = G(t) [\bm{X}(t-\tau) - \bm{X}(t)] \:,
\end{eqnarray}
where $\bm{X}(t)$ and $\bm{X}(t-\tau)$ are the current state vector and the time delayed state vector, with delay period $\tau,$ \modTY{and} $G(t)$ is the control gain, which in general could be some matrix (thereby coupling feedback between degrees of freedom) but here is a function of time only. 
Note that when a time periodic state with period $\tau,$ or a time-independent state, is stabilised successfully, the control force (\ref{eq:pyragas2}) will decay exponentially towards zero (\modDL{assuming $G(t)$ remains bounded}). This can only occur if the stabilised state is itself a solution of the uncontrolled system, so that TDF is considered to be a non-invasive control method.
One reason for the lack of application of TDF in fluid systems is likely due to the so-called ``odd-number'' limitation. 
This claims that states with an odd-number of unstable Floquet multipliers are unable to be stabilised by this method \citep{Just1999,  Nakajima:1998kw}.
\citet{Nakajima1997} explains the issue from a bifurcation perspective; the non-invasive feature means that the number of solutions of period $\tau$ cannot vary with $G,$ however any stabilisation requires a change of stability and hence bifurcation. Such a bifurcation cannot, therefore, be of pitchfork or saddle-node type (changing $\tau$-period solutions), and so must involve the crossing of a complex conjugate pair of exponents in a Hopf or period-doubling bifurcation. There have been numerous studies offering resolutions to this issue, including forcing oscillation of the unstable manifold through $G$ \citep{Schuster1997,Flunkert:2011cg}, an `act-and-wait' approach \citep{pyragas_2019_state-dependent}, \modDL{using time-dependent gains more generically \citep{Sieber:2016cu}}, using symmetries \citep{Nakajima_1998,lucas_2022_stabilization} and \modDL{a notable} counter example \modDL{where a transcritical bifurcation can be stabilised under certain conditions} \citep{Fiedler:2011db}. 

\citet{Shaabani2017} report the application of Pyragas control to suppress vortex pairing in a periodically forced jet. This work approaches the control method as a frequency damping technique; filtering out non-harmonic frequencies, leaving only $\tau$ behind. Here the odd-number issue can be viewed as a zero-mode limit where there is no incipient frequency to damp. 
Recently, \citet{lucas_2022_stabilization} applied this method 
to two-dimensional Kolmogorov flow turbulence, validating the stabilisation of the base flow via linear stability analysis and showing successful stabilisation of several equilibria and travelling waves. This was achieved by including the symmetries of the target solutions into the control force. In many examples, including the stabilisation of the laminar solution, this was an effective means to avoid the odd-number problem (not too dissimilar to, but different from, the ``half-period'' approach in \cite{Nakajima_1998}). An adaptive, gradient descent, approach is also used to obtain the relative translations of travelling waves so that the method can be successful without any foreknowledge of the ECS.
%
%
%
%

%
%
%
Our main objective in this paper is to present an improved method for using TDF to stabilise nonlinear traveling waves in pipe turbulence, where the flows are more physically relevant, 
more unstable and have more spatial complexity. 

%
The vast majority of TDF research has been devoted to investigating the linear stability of a target solution, it being a necessary property for successful stabilisation \citep{Nakajima1997}. 
However, this is not a sufficient condition as successful practical stabilisation can also depend on the initial conditions. In high dimensional systems it would be advantageous to design the control to also maximise the basin of attraction of the stabilised state. 
Furthermore it will be necessary for the TDF control to stabilise several unstable directions, each with differing eigenfrequencies, without knowing what these frequencies are before initiating the control.
To achieve this, we develop an adaptive version of the 
multi-frequency damping TDF control method \citep{akervik_2006_steady,Shaabani2017}, 
i.e., multiple time-delayed feedback (MTDF), where several TDF terms are used.
One drawback of the MTDF approach is the need to optimise a separate control gain for each delay period applied. In order to avoid a trial-and-error sampling of this high-dimensional parameter space, we apply a gradient descent method \citep{Lehnert:2011hu} to evolve each $G$ towards their stabilising values.

This paper is organised as follows. 
In \S~2, we describe the governing equations for pipe flow with the control force, 
the numerical method, the continuous and discrete symmetries, 
and define relevant flow measures. 
In \S~3, after introducing single time-delayed feedback for pipe flow 
and the adaptive translation method, we demonstrate some successful cases 
stabilising an unstable travelling wave at low Reynolds number \modDL{and in certain symmetry subspaces}. 
We predict the behaviour of TDF by using an approximate linear stability analysis and control theory, 
in such a way identifying the optimal control parameters for this solution. 
In \S~4, 
we introduce multiple delayed feedback control (MTDF) and demonstrate
its effectiveness in stabilising more highly unstable states and analyse its behaviour from the frequency damping perspective. 
%
In \S~5, 
we demonstrate 
successful stabilisation of travelling waves from relatively high Reynolds number turbulence, using MTDF alongside the optimisation methods for gains and translations.  \modDL{In the course of doing this, we demonstrate the stabilsation of two unknown solutions from highly symmetric subspaces.}
%
%

\section{Numerical formulation}\label{sec:simulation}

\subsection{Pipe flow with time-delayed feedback}
\label{sec:equations}
In this study, we consider the incompressible, viscous flow 
in straight, \modDL{cylindrical}, pressure-driven pipes. 

\subsection{Direct numerical simulations}\label{dns}
\modTY{
In this study, we perform direct numerical simulations of pipe flow 
with the time-delayed feedback `force' being applied. 
}
We solve the system outlined in \S~\ref{sec:equations} numerically 
using the open-source code \texttt{openpipeflow} \citep{willis_2017_the} 
which allows the relatively easy implementation of our control terms.
Our computational domain is periodic in both azimuthal and streamwise directions,
and the streamwise length of the periodic pipe is $L=2\pi/\alpha$, e.g., 
with $\alpha=1.25$ corresponding to \modTY{$L\approx 5 R$}. 
Here, $\bm{u}$ and \modTY{$\hat{p}$} are both expanded in discrete Fourier series 
in the streamwise and azimuthal directions.
Spatial derivatives with respect to $r$ 
are evaluated based on a 9-point finite differences stencil with a non-uniform mesh, 
for which 1st/2nd order derivatives are 
calculated to 8th/7th order \citep{willis_2017_the}.
With these spatial discretisation schemes, 
$\bm{u}$ is expressed as 
\begin{eqnarray}\label{eq:}
\bm{u}(r_n,\theta,z,t) = \sum_{|k|<K} \sum_{|m|<M} 
\widetilde{\bm{u}}_{km}(r_n,t) \mathrm{e}^{i(\alpha k z+ m \theta)} 
\end{eqnarray}
where $\widetilde{\bm{u}}_{km}$ are the Fourier coefficients of $\bm{u},$ $r_n$ ($n=1 \ldots N$) denotes the radial grid points (non-uniformly distributed on [0, 1]), and $k$ and $m$ are the streamwise and azimuthal wavenumbers respectively, with $K$ and $M$ the de-aliasing cutoff \citep[see][]{willis_2017_the}. 
%
%
%
The resolution of a given calculation is described by a vector $(N,M,K)$ and is adjusted until the energy in the spectral coefficients drops by at least 5 and usually 6 decades. 
%
Time-stepping is via a second-order predictor-corrector scheme, 
with Euler predictor for the non-linear terms and Crank-Nicolson for the viscous diffusion.
%
%

\subsection{Continuous and discrete symmetries}\label{sym}
%
The equations of pipe flow are invariant under continuous translations in $z$
\begin{eqnarray}\label{eq:sym_cont_z}
\mathcal{T}_z(s_z) 
[u_r,u_\theta,u_z,p](r,\theta,z) \rightarrow 
[u_r,u_\theta,u_z,p](r,\theta,z+s_z) \:, 
\end{eqnarray}
continuous rotations in $\theta,$
\begin{eqnarray}\label{eq:sym_cont_th}
\mathcal{T}_\theta(s_\theta)  
[u_r,u_\theta,u_z,p](r,\theta,z) \rightarrow 
[u_r,u_\theta,u_z,p](r,\theta+s_\theta,z) \:, 
\end{eqnarray}
and reflections about $\theta=0$
\begin{eqnarray}\label{eq:sym_ref}
\sigma[u_r,u_\theta,u_z,p](r,\theta,z) \rightarrow 
[u_r,-u_\theta,u_z,p](r,-\theta,z) \:, 
\end{eqnarray}
%
%
%
%

Following the approach by \citet{Willis:2013bu}, 
we will restrict our investigations to dynamics restricted 
to the `shift-and-reflect' symmetry subspace, $S=\sigma \mathcal{T}_z\left (\frac{L}{2}\right )$, 
\begin{eqnarray}\label{eq:subspace_S}
S[u_r,u_\theta,u_z,p](r,\theta,z) \rightarrow [u_r,-u_\theta,u_z,p](r,-\theta,z-L_z/2) \:, 
\end{eqnarray}
and the discrete `rotate-and-reflect' symmetry $Z_{m_p}=\sigma\mathcal{T}_\theta \left (\frac{\pi}{m_p} \right)$, 
\begin{eqnarray}\label{eq:subspace_Z}
Z_{m_p}[u_r,u_\theta,u_z,p](r,\theta,z) 
\rightarrow [u_r,-u_\theta,u_z,p](r,\pi/m_p-\theta,z) \:.
\end{eqnarray}
%
%
%
%
%
Note $m_p=1$ denotes azimuthal periodicity (the full space), and  
$m_p$-fold rotational symmetry is enforced for $m_p\ge 2$ 
\citep{wedin_2004_exact,pringle_2009_highly,Willis:2013bu}. 
%
%
By imposing $S$ the flow is `pinned' in $\theta$ and continuous rotations are prohibited. This means that our solutions are only permitted to travel in the streamwise direction due to $\mathcal{T}_z.$
\subsection{Flow measures}\label{diagnostic}
In order to monitor the system behaviour, 
we consider the spatially integrated quantities from the energy budget equation 
\begin{eqnarray}\label{eq:energybudget}
\frac{\mathrm{d} E}{\mathrm{d} t} = I - D + I_{\mathrm{TDF}} \:, 
\end{eqnarray}
where $E$ is the total kinetic energy, $I$ is the total energy input due to the imposed pressure gradient and  $D$ is the total energy dissipation
\modTY{
\begin{eqnarray}\label{eq:energy}
E \coloneqq \frac{1}{2} \int |\bm{U}|^2 \mathrm{d}V \:,\,
I \coloneqq \int \bm{U}\cdot(-\nabla P)\mathrm{d}V \:,  \,
D \coloneqq \frac{1}{Re} \int | \nabla \times \bm{U} |^2 \mathrm{d}V \:, 
\end{eqnarray}
}
and $I_{\mathrm{TDF}}$ is the total energy input due to feedback force $\bm{F}$,  
\begin{eqnarray}\label{eq:input}
I_{\mathrm{TDF}} \coloneqq \int \bm{U}\cdot\bm{F} \mathrm{d}V \:.
\end{eqnarray}
For the laminar state, 
$I_{\mathrm{TDF}}=0$ and the values of energy, input rate and dissipation rate $E_{\mathrm{lam}}$, $I_{\mathrm{lam}}$, and $D_{\mathrm{lam}}$ can be computed \modDL{from equations \eqref{eq:energy} with $\bm{U}=\bm{U}_{\mathrm{HP}}.$} 
%
%

\section{Time-delayed feedback}\label{SS:STDF}
\subsection{Formulation}\label{ss:formulation}
%
In this section 
we outline the application of the time-delayed feedback control (\ref{eq:pyragas2}) 
to the pipe flow.
In order to effectively stabilise travelling wave solutions, 
the delayed state must be translated by a streamwise shift $s_z$ in relation to the phase speed, 
$c_z$, such that $c_z=\frac{s_z}{\tau}$ for a given time delay $\tau.$
Using the translation operator (\ref{eq:sym_cont_z}), 
the most basic form of TDF force in pipe flow may be formulated as
\begin{eqnarray}\label{eq:STDF}
\bm{F}_{\mathrm{TDF}}(r,\theta,z,t) = 
G(t) [  \mathcal{T}_z(s_z) 
\bm{u}(r,\theta,z,t-\tau) - \bm{u}(r,\theta,z,t)] \:, 
\end{eqnarray}
where $\tau$ is the delay period, and $G(t)$ is the gain. 
Note the gain here is a simple scalar function of time, but it may itself be an operator (matrix) or spatially inhomogenous. In the full space one may also wish to apply a rotation $\mathcal{T}_{\theta}.$ 
%

In order to avoid a discontinuity propagating through the solution when the control is initiated, 
we set $G(t)$ using the following sigmoid function: 
\begin{eqnarray}
\label{eq:sigmoidgain}
G(t) = \begin{cases}
\frac{G^{\mathrm{max}}}{1+\exp[a(b+t_s-t)]} \:, & \qquad t>t_s\\
  0 & \qquad t\leq t_s.
\end{cases}
\end{eqnarray}
Here, $t_s$ is start time (the time TDF is initiated), 
$G^{\mathrm{max}}$ is the maximum gain, 
and $a$ determines the slope of $G(t)$ and 
the half-height time, $t_h=t_s+b,$ set by $b,$ is
when $G(t_h)=G^{\mathrm{max}}/2$. 
%

The value of the translation, $s_z,$ 
for successful stabilisation will be, in general, unknown in advance, 
so we require a method to evolve $s_z$ towards the required value.
We implement the adaptive method developed in \citet{lucas_2022_stabilization}. 
This method allows $s_z$ to vary via gradient descent by solving 
an ordinary differential equation 
\begin{eqnarray}
\label{eq:ode_shift}
\frac{\mathrm{d} s_z}{\mathrm{d} t} = \gamma_s \delta s_z,  
\end{eqnarray}
where $\gamma_s$ is a parameter controlling the speed of the descent 
and $\delta s_z$ is, near a travelling state, an estimate of 
the streamwise translation remaining between 
the current state and the delayed and translated state. 
Specifically, $\delta s_z$ is computed as 
\begin{eqnarray}\label{eq:ds_z}
\delta s_z = s_z^{\mathrm{est}} - s_z \:, 
\end{eqnarray}
where $s_z^{\mathrm{est}}$ is a streamwise translation 
that is dynamically estimated. 
Here, $s_z^{\mathrm{est}}$ can be computed 
via the time series of $c_z(t)$ such that 
\begin{eqnarray}\label{eq:sz_est}
s_z^{\mathrm{est}}(t;\tau) = \int_{t-\tau}^{t} c_z(t^{\prime}) \mathrm{d} t^{\prime} \:. 
\end{eqnarray}
We can compute $c_z(t)$ using complex phase rotations such that 
\begin{eqnarray}\label{eq:c_z}
c_z(t) = 
\frac{1}{N_{\mathrm{e}}} \sum_{k= \pm 1}\sum_{m= \pm 1}\sum_{n=1}^{N-1} 
\frac{1}{i \alpha k \Delta t} \arg 
\bigg[ \frac{\widetilde{u}_{k m}(r_n, t)}{\widetilde{u}_{k m}(r_n, t-\Delta t)} \bigg] 
\end{eqnarray}
where $N_{\mathrm{e}}=4(N-1)$ is 
the number of the elements in the above summation 
and the wall points ($n=N$) are excluded.  
Note that $c_z(t)$ is considered an average phase speed 
over one time step, $\Delta t$, in the streamwise direction. 
%
%
We solve the ODE (\ref{eq:ode_shift}) while computing $s_z^{\mathrm{est}}$ 
using (\ref{eq:sz_est}) alongside the DNS 
with a second-order Adams-Bashforth time stepping scheme. 
As will be seen in the later sections, using long time delays can be essential 
in successful stabilisation of unstable travelling waves in a straight \modDL{cylindrical} pipe. 
Computing estimated translations using (\ref{eq:sz_est}) is slightly different 
from the original version in \citet[][]{lucas_2022_stabilization}. 
The current version is found to perform better than the original when encountering long delays. 

\begin{table}
\begin{center}
\begin{tabular}{lccccccccccc}
sol. & $Re$ & $\alpha$ & $E/E_{\mathrm{lam}}$ & $I/I_{\mathrm{lam}}$  & $Re_{\tau}$ & $c_z$ 
&  sym. & $\mu_i>0$ & $\textrm{max}(\mu_i)$ & $\textrm{min}(\omega_i)$ s.t. $\mu_i>0$ \\
{\textsf{ML}}  & $2400$ & 1.25 & 0.88662 & 1.6970 & 90.3 & 0.71049 & $S$,$Z_2$ & 1c     & 0.0061985 & 0.018295 \\
               &        &      &         &        &      &         & $S$       & 1r+3c  & 0.0676    & 0.018295 \\
{\textsf{UB}}  & $2400$ & 1.25 & 0.85273 & 2.5102 & 110  & 0.64924 & $S$,$Z_2$ & 3c     & 0.056285  & 0.32393  \\
               &        &      &         &        &      &         & $S$       & 9c     & 0.10689   & 0.048437 \\
               &        &      &         &        &      &         & --        & 2r+17c & 0.10751   & 0.048437 \\
               & $3700$ & 1.25 & 0.84822 & 2.4866 & 136  & 0.65056 & $S$,$Z_2$ & 6c     & 0.077546  & 0.19076  \\ 
\textsf{S2U}   & $2400$ & 1.25 & 0.89383 & 1.4695 & 84.0 & 0.64763 & $S$       & 1c     & 0.014152  & 0.13040  \\
               &        &      &         &        &      &         & --        & 1r+2c  & 0.014152  & 0.12385  \\
\textsf{N3}    & $2400$ & 2.5  & 0.83421 & 3.0283 & 121  & 0.59508 & $S$,$Z_3$ & --     & --        & --       \\
               &        &      &         &        &      &         & $S$       & 1r+6c  & 0.091251  & 0.084125 \\
               &        &      &         &        &      &         & --        & 1r+11c & 0.091250  & 0.084125 \\
{\textsf{N4U}} & $2500$ & 1.7  & 0.84911 & 3.2793 & 128  & 0.52575 & $S$,$Z_4$ & 4c     & 0.24623   & 0.10674  \\ 
               &        &      &         &        &      &         & $S$       & 14c    & 0.24623   & 0.047864 \\ 
               &        &      &         &        &      &         & --        & 1r+28c & 0.24623   & $9.9240 \times 10^{-4}$\\
{\textsf{N5}}  & $2500$ & 2    & 0.88591 & 3.0155 & 123  & 0.47483 & $S$,$Z_5$ & 3c     & 0.1999    & 0.0324   \\ 
{\textsf{N7}}  & $3500$ & 3    & 0.95849 & 3.2916 & 152  & 0.39523 & $S$,$Z_7$ &    2c  &   0.1775  & 0.0821  
\end{tabular}
\caption{
Table summarising the properties of all travelling waves studied in this paper. 
$Re = R U_{\mathrm{cl}}/\nu$ is the Reynolds number, 
$\alpha$ is the streamwise wavenumber \modDL{($L=2\pi/\alpha$)}, 
$E$ is the total kinetic energy, 
$I$ is the total energy input rate, 
$Re_{\tau}=\sqrt{2 Re I/I_{\mathrm{lam}}}$ is the friction Reynolds number, 
$c_z$ is the streamwise phase velocity,  \modDL{``sym.'' refers to the symmetry subspace}, 
$\lambda_i = \mu_i \pm \mathrm{i} \omega_i$ 
are the eigenvalues of the solutions \modDL{such that the final three columns are respectively the number and type of unstable eigenvalues, the most unstable growth rate and the smallest unstable eigenfrequency.}
%
}
\label{tab:1}
\end{center}
\end{table}

\subsection{Validation -- Stabilising weakly unstable solutions}\label{SS:STDF_validation}
\begin{figure}
\centering
\includegraphics[width=\linewidth,clip]
{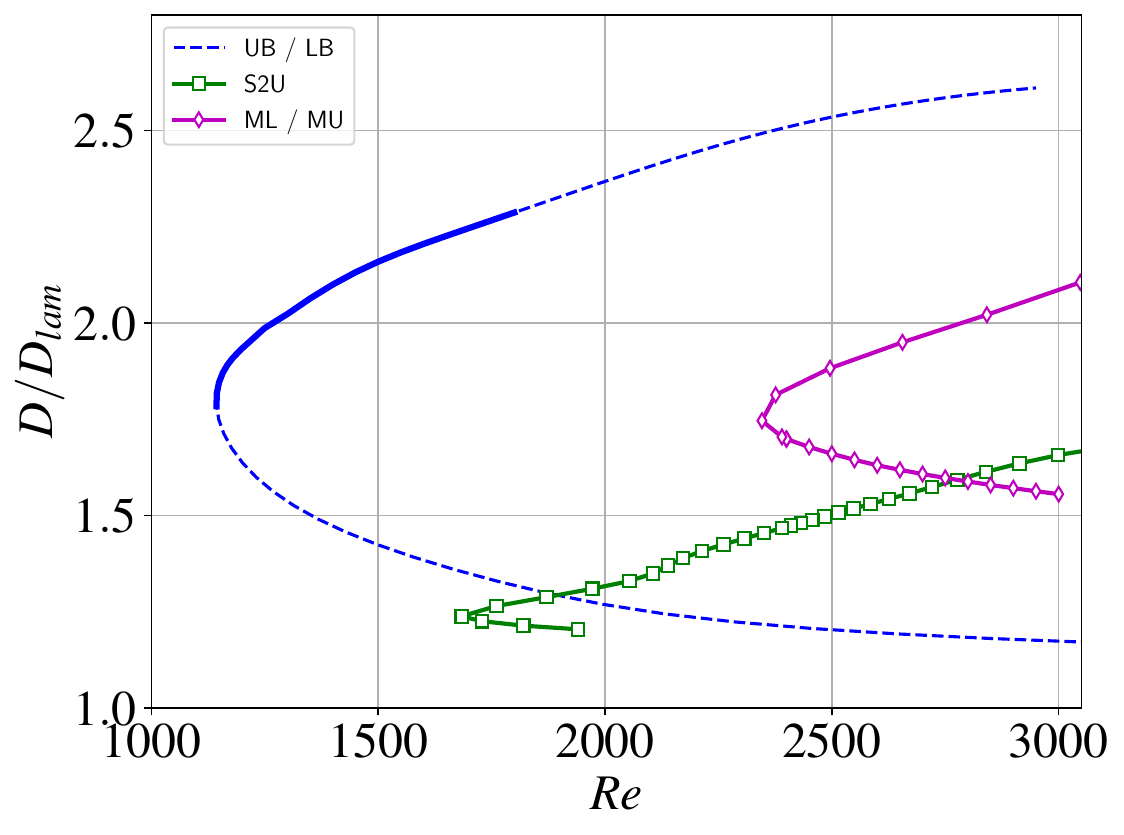}
\caption{
Bifurcation diagram showing solution branches for flows with $\alpha=1.25$.  
}
\label{fig:bif}
\end{figure}

\begin{figure}
\centering
\subfigure{
\begin{minipage}{0.5\linewidth}
\centerline{(a)}
\includegraphics[width=\linewidth,clip]{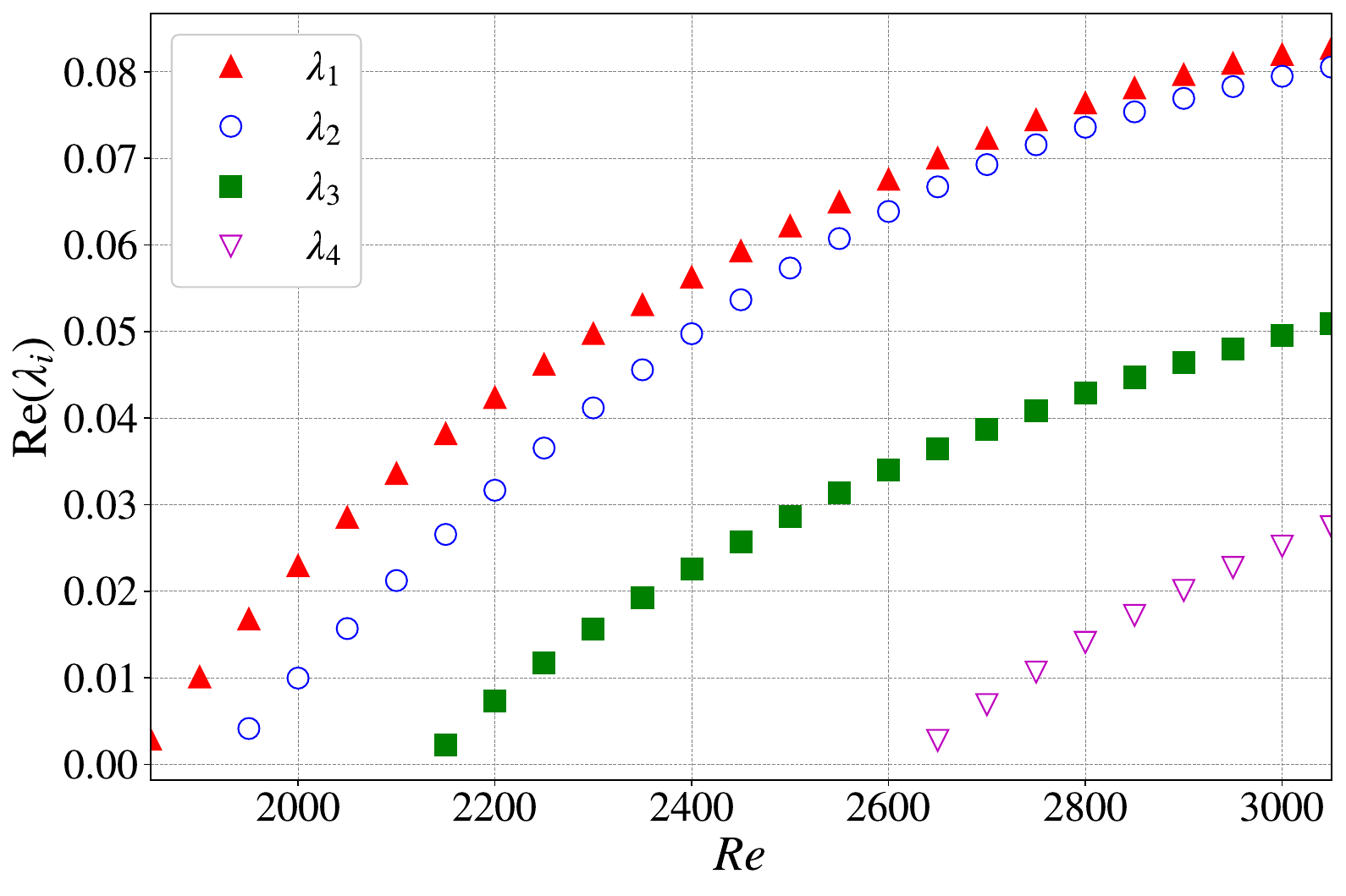}
\end{minipage}
\begin{minipage}{0.5\linewidth}
\centerline{(b)}
\includegraphics[width=\linewidth,clip]{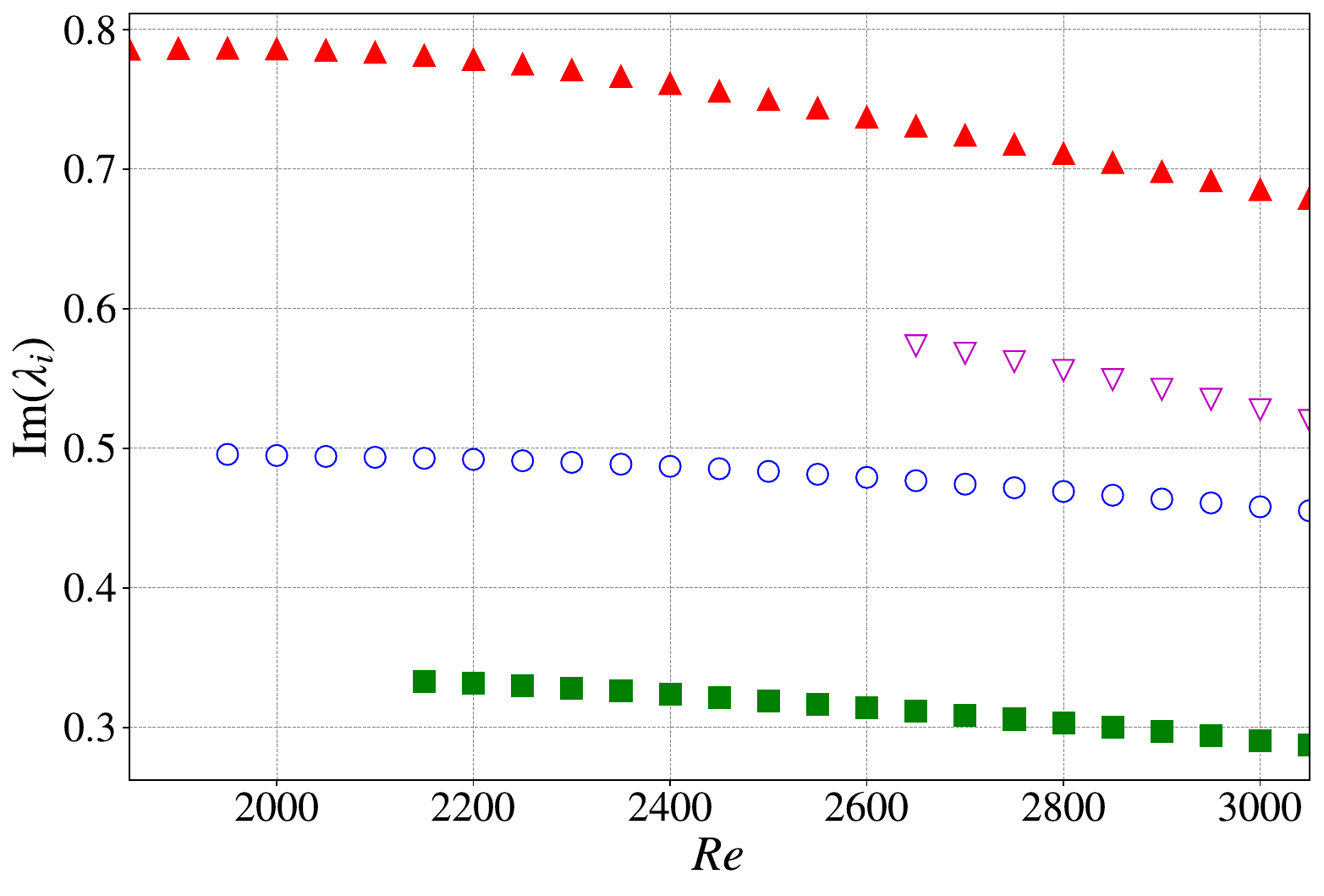}
\end{minipage}
}
\caption{
The $Re$-dependence of leading eigenvalues of the \textsf{UB} solution. 
(a) Real part and (b) imaginary part of unstable complex eigenvalues. 
No purely real eigenvalue exists; thus the ``odd-number limitation'' is not encountered.  
}
\label{fig:eigenvalues}
\end{figure}

\begin{figure}
\centering

\subfigure{
\begin{minipage}{0.5\linewidth}
\centerline{(a)}
\includegraphics[width=\linewidth,clip]{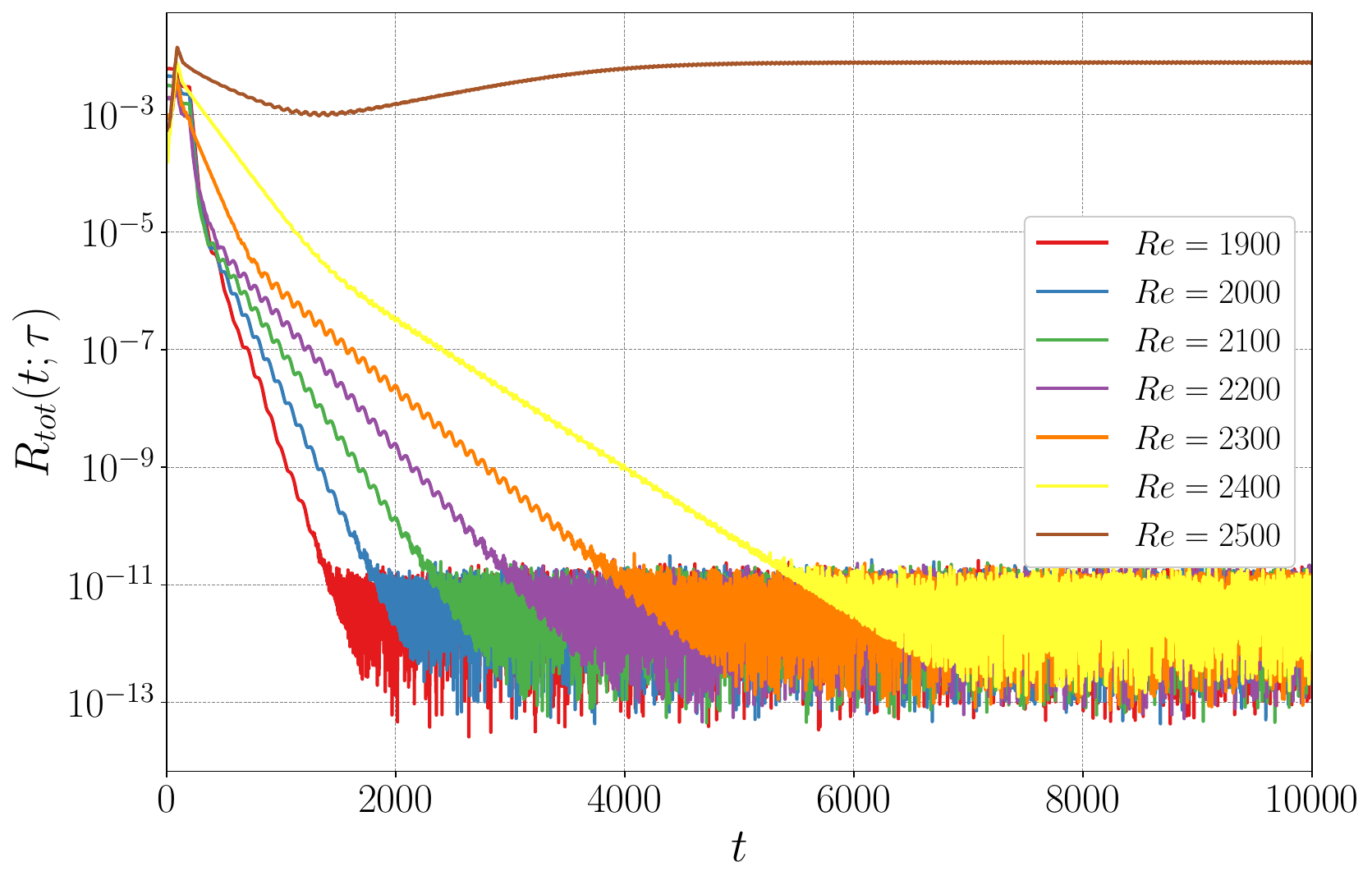}
\end{minipage}
\begin{minipage}{0.5\linewidth}
\centerline{(b)}
\includegraphics[width=\linewidth,clip]{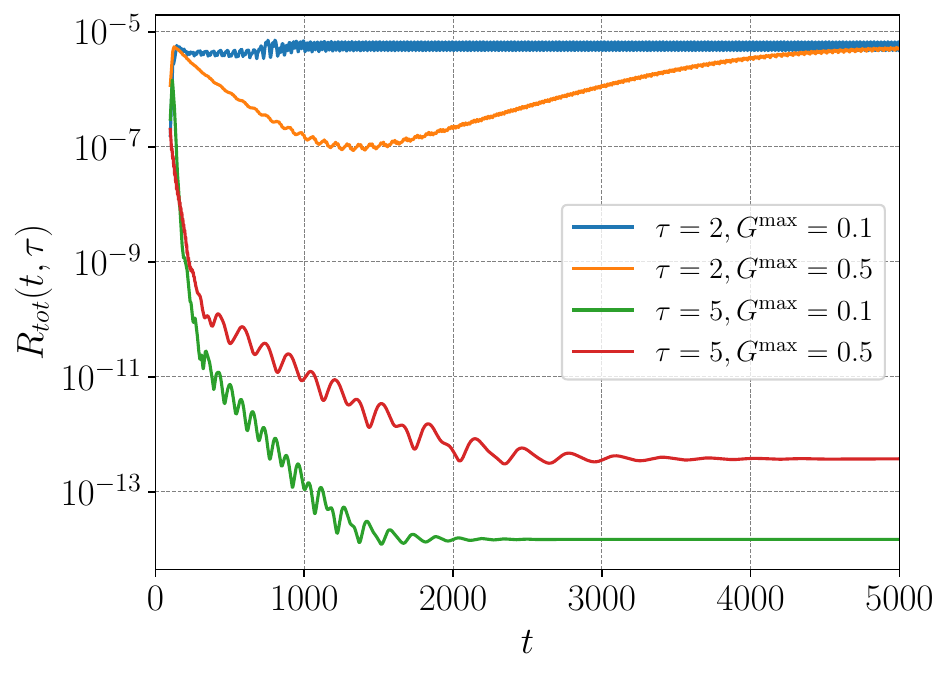}
\end{minipage}
}
\caption{
Time series in the relative residual, $R_{\mathrm{tot}}(t;\tau)$ for attempted TDF stabilisation of \textsf{UB}. 
(a) shows $Re=1900$,$2000$,$2100$,$2200$,$2300$,$2400$,$2500$ for TDF parameters $\tau = 2$ and $G^{\max}=0.5,$ stabilisation fails for $Re>2400$. Right plot (b) shows the $Re=2500$ case with $G^{\max}=0.1$ and 0.5, and $\tau=2$ \modDL{(both of which fail to stabilise)} and  $\tau =5$ \modDL{(which stabilise successfully)}, demonstrating consistency with the linear analysis. The simulations are initiated with \textsf{UB} at each Reynolds number. {Note that recording $R_{\mathrm{tot}}(t;\tau)$ initiates at $t=\tau$.}
 }
\label{fig:Q}
\end{figure}

%
%
In the absence of any simpler solutions to use for validation (the laminar HP flow is linearly stable at least up to $Re=10^7$, see \citet{meseguer_2003_linearized}), we attempt to stabilise known unstable travelling waves \citep{Willis:2013bu} at low $Re$ first. 
For the parameter values of $\alpha=1.25$ ($L_z \approx 5R$) and $Re=2400$, we are able to stabilise the travelling waves \textsf{ML}, \textsf{UB}, \textsf{S2U} in their respective symmetric subspace (see table~\ref{tab:1}). 
These travelling waves have only complex unstable eigenvalues 
at $\alpha=1.25$ and $Re=2400$ and in their symmetry subspaces, meaning that the odd-number issue is not applicable, or rather is avoided by projection into those subspaces 
\citep[see][for a discussion]{lucas_2022_stabilization}.
%

In order to investigate the effectiveness of TDF as the stability of a target solution changes, we concentrate on the \textsf{UB} solution \modDL{(in $(S,Z_2)$)} and vary $Re,$ returning to the other solutions in \S~\ref{sec:TWs}.
Figure~\ref{fig:bif} shows the bifurcation diagram with the continuation of the \textsf{UB} solution branch (blue) as a function of $Re$ (alongside \textsf{ML} and \textsf{S2U}). This is generated numerically using the Newton--GMRES--hookstep method \citep{Viswanath:2007wc,Viswanath:2009vu} packaged with the \texttt{openpipeflow} code \citep{willis_2017_the}. 
This confirms that \textsf{UB} originates in a saddle-node bifurcation at $Re\approx1150$. On increasing $Re$, the stable upper branch (solid line in figure \ref{fig:bif}) of \textsf{UB} becomes unstable via a supercritical Hopf bifurcation at $Re\approx1800$ (unstable solution in dashed blue). 

In figure~\ref{fig:eigenvalues} we plot the real and imaginary parts of the complex growth rates of the \textsf{UB} solution which demonstrates the expected increase in instability \modDL{as the Reynolds number increases}. A sequence of further bifurcations are observed and new unstable directions  are formed as more eigenvalues cross the \modDL{imaginary} axis. The solution has only complex eigenvalues at least up to $Re=3050$, therefore it is free from the odd-number limitation \citep{Nakajima1997}. 

\modDL{Using} this stability information we attempt to stabilise \textsf{UB} at various $Re$ using TDF. 
The initial condition is \textsf{UB} such that $\bm{U}(r,\theta,z,0)=\bm{U}_{\mathsf{UB}}$. 
We set $t_s=10$ such that the trajectory is still close to the \textsf{UB} solution when TDF is activated, avoiding the need to consider if our initial condition falls inside the basin of attraction for now. 
In this sense this section verifies a necessary condition for stabilisation and we will investigate more generic initial conditions later. \modDL{In all cases where \textsf{UB} is stabilised an initial $c_z(0)=0.65$ is used.}
%

In order to quantify the size of the TDF force term, 
we define the relative residual, $R_{\mathrm{tot}}$, 
between the current state and the translated and delayed state, as 
\begin{eqnarray}\label{eq:residual}
R_{\mathrm{tot}}
= \frac{\|\mathcal{T}_z(s_z)\bm{U}(r,\theta,z,t-\tau)-\bm{U}(r,\theta,z,t)\|_2}
{\|\bm{U}(r,\theta,z,t)\|_2}
\end{eqnarray}
where $ \| \bm{A} \|_2 = \sqrt{\frac{1}{2} \langle \bm{A}\cdot\bm{A} \rangle_V.}$ 
%
Figure~\ref{fig:Q}(a) shows time series in $R_{\mathrm{tot}}$ 
for seven Reynolds numbers, $Re=1900$, 2000, 2100, 2200, 2300, 2400, 2500, 
with $\tau=2,$ $G^\mathrm{max}=0.5,$ 
$\gamma_s=0.1,$ $t_s=10,$ $a=0.1,$ $b=100.$ 
At $Re=1900$, stabilisation of \textsf{UB} is achieved quickly 
such that $R_{\mathrm{tot}}$ has decreased to $O(10^{-12})$ by $t=2000$. 
The rate of attraction of the travelling wave decreases 
as $Re$ increases with all other parameters held fixed, \modDL{as one might expect}.

At $Re=2500$, stabilisation is unsuccessful; even by increasing $G^\mathrm{max}$ 
we are unable to stabilise the \textsf{UB} state at this $Re,$ \modDL{without adjusting the other TDF paramters}.
It turns out that the growth rate is not the key quantity preventing stabilisation in this case, 
and neither is it the appearance of an odd-numbered eigenvalue 
(in contrast to the findings in \cite{lucas_2022_stabilization}).
\textsf{UB} at $Re=2500$ has three unstable eigenfrequencies, 
which are decreasing as a function of $Re;$ it is these values which 
fall outside the domain of stabilisation of our method. 
In the next subsection, 
we analyse the TDF control from the perspective of ``frequency damping''  
\citep{akervik_2006_steady,Shaabani2017} and linear stability analysis.

\subsection{Frequency damping and linear stability with TDF}\label{sec:linear}

\begin{figure}
\centering
\subfigure{
\begin{minipage}{0.5\linewidth}
\centerline{(a)}
\includegraphics[width=\linewidth]{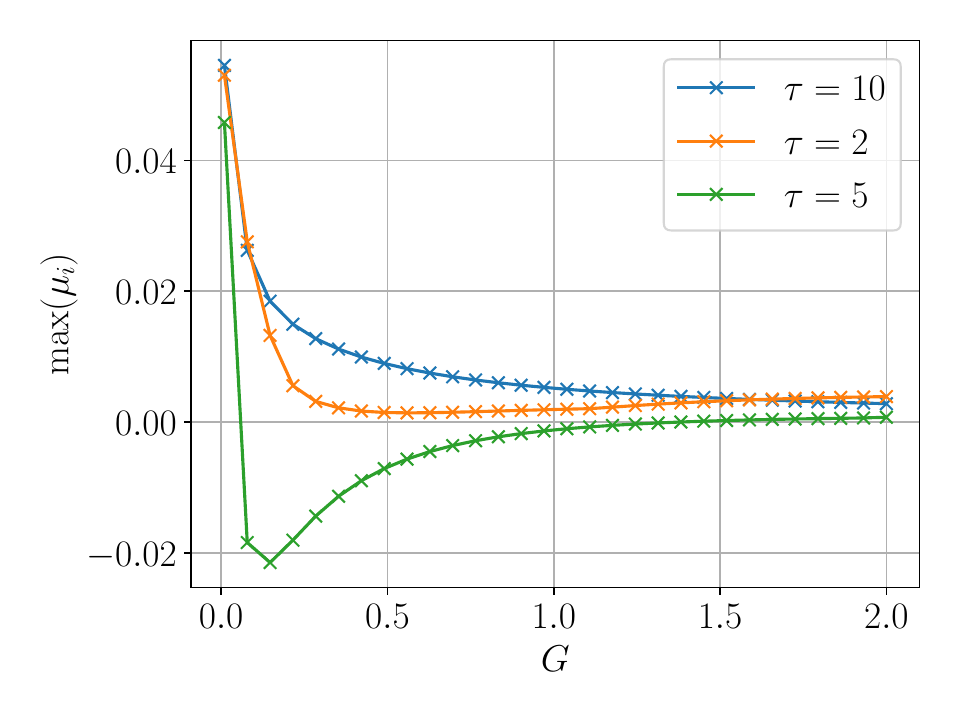}
\end{minipage}
\begin{minipage}{0.5\linewidth}
\centerline{(b)}
\includegraphics[width=\linewidth]{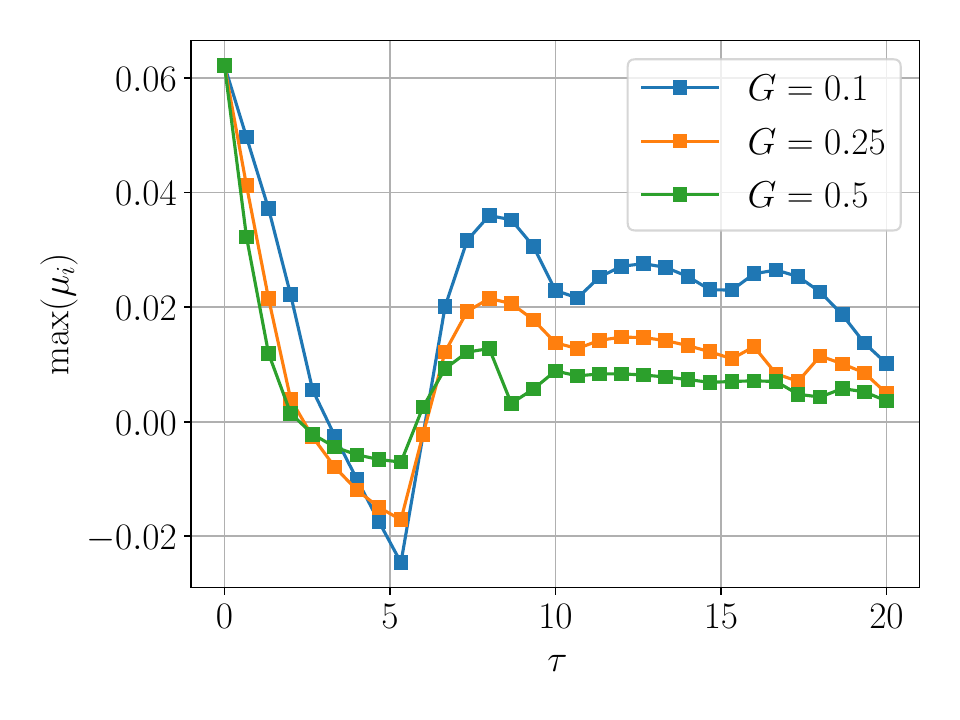}
\end{minipage}
}
\caption{Dependence of the largest real-part of the eigenvalue spectrum $\max_i \mu_i$  ($\lambda_i = \mu_i + \mathrm{i} \omega_i,$) with $G$ (left (a)) and $\tau,$ (right (b)) for $Re=2500$ \textsf{UB} solution, according to the approximate linear theory.}
\label{fig:TDF_LambdaMax}
\end{figure}

Predicting the outcome of TDF in these scenarios is difficult due to the relatively high dimension of the unstable manifold (of the uncontrolled solution) and the highly nonlinear nature of the solution. Nevertheless we can approximate the effect of the TDF terms on the unstable part of the eigenvalue spectrum. Starting from equation \eqref{eq:pyragas1}-\eqref{eq:pyragas2}, assuming $\bm{X} = \bar{\bm{X}} + \bm{v}e^{\lambda t},$ with $\bar{\bm{X}}$ a steady state solution to \eqref{eq:pyragas1}, assuming $G$ is now constant and linearising in the usual way results in the modified eigenvalue problem

\begin{equation}
   \lambda \bm{v} = \mathcal{J}\bm{v} + G\bm{v}\left(e^{-\lambda \tau} - 1\right), \label{eq:linear}
\end{equation}
with eigenvalue $\lambda,$ eigenvector $\bm{v}$ and $\mathcal{J}$ being the Jacobian of $\bm{f}.$ The eigenvalue problem is now transcendental and so requires some numerical root searching to obtain the eigenvalues. \cite{lucas_2022_stabilization} tackled this difficulty by assuming $|\lambda \tau| \ll 1$ and expanding the exponential, however we are unable to make this assumption here. Moreover, root-searching for $\lambda$ in the full problem, i.e. some numerical linearisation of (\ref{eq:NSmom}), would be impractical. Instead we create a ``synthetic'' $\mathcal{J},$ which we design to have the same unstable eigenvalues as the uncontrolled solution, and some randomly chosen eigenvectors. \modDL{For instance,} at $Re=2500$, with three pairs of unstable, complex eigenvalues, this amounts to a 6 dimensional real-valued system. We can then solve the transcendental characteristic equation numerically with a simple Newton search to obtain all of the roots and hence eigenvalues. This procedure is able to predict well the effect of TDF on the \textsf{UB} solution; \modDL{across} the parameters shown in figure \ref{fig:Q}(a), \modDL{and using the uncontrolled eigenvalues shown in figure \ref{fig:eigenvalues},} the analysis finds that the largest eigenvalue crosses the imaginary axis at $Re \approx 2450.$ Moreover \modDL{the method} is sufficiently efficient that we can explore the $(G,\tau)$ parameter space. Figure \ref{fig:TDF_LambdaMax} shows two plots of the largest real-part of the eigenvalues of our approximated TDF system at $Re=2500$. We observe that for $\tau = 2$ no value of $G$ results in stability, however increasing slightly to $\tau=5$ results in a region of stability for relatively small $G,$ we see $G\approx 0.1$ is roughly optimal. This is slightly surprising as one would have assumed that as solutions become more unstable the remedy would be to apply larger gains to move eigenvalues back across the imaginary axis. It should be noted that this analysis is only an approximation to the problem, in particular it does not predict whether TDF may destabilise stable modes \modDL{(while it is possible for TDF to destabilise stable modes, or introduce new modes of instability, as found in \cite{linkmann_2020}, this is never observed when the uncontrolled modes are also stabilised and only for large $G$ and $\tau.$). }

\begin{figure}
\centering
\begin{minipage}{0.47\linewidth}
\centerline{(a)}
\includegraphics[width=\linewidth]{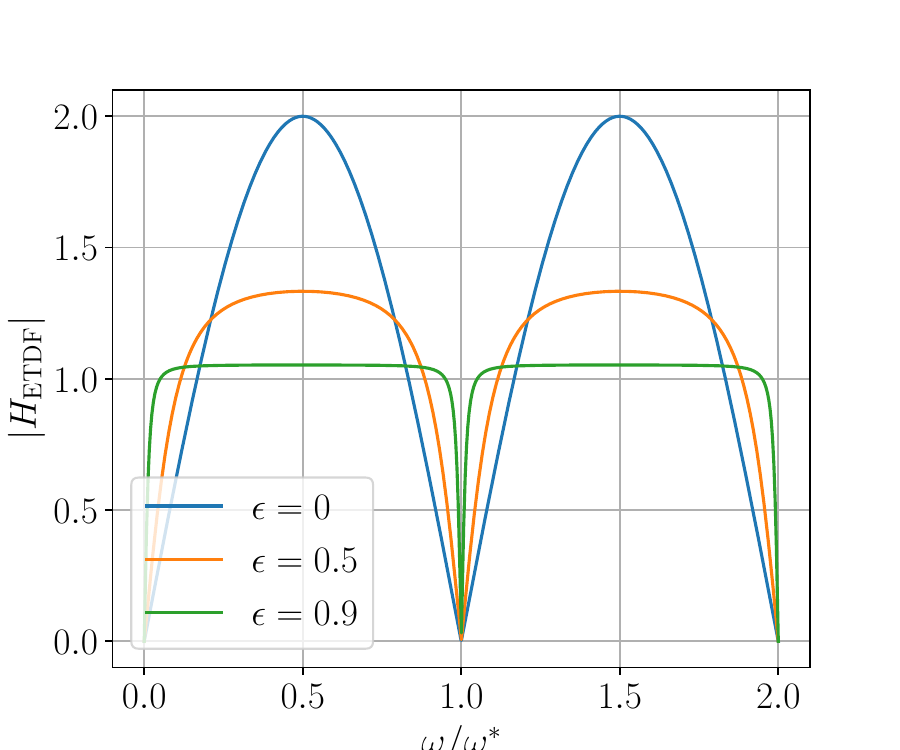}
\end{minipage}
\begin{minipage}{0.47\linewidth}
\centerline{(b)}
\includegraphics[width=\linewidth]{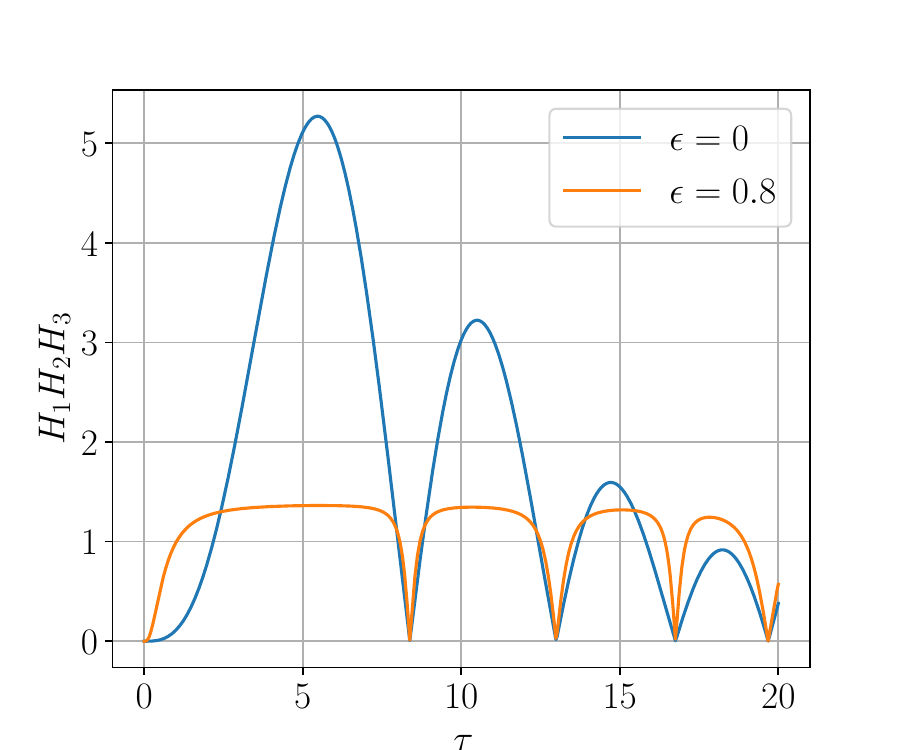}
\end{minipage}
\caption{Transfer functions for TDF ($\epsilon = 0$) and ETDF (various $\epsilon\neq 0$). 
(a) shows $H_{\mathrm{ETDF}}(\omega/\omega^*)$ where $\omega^* = 2\pi/\tau$ demonstrating 
peaks at subharmonics and zeros at harmonics of $\tau.$ 
(b) shows the product 
of $H_n = H_{\mathrm{ETDF}}(\omega_n,\tau)$ 
for the three unstable eigenfrequencies of \textsf{UB} at $Re=2500.$ 
This indicates an optimal $\tau,$ for TDF, of around 5.5. 
This is consistent with the linear analysis 
shown in figure~\ref{fig:TDF_LambdaMax}. 
}
\label{fig:prodH}
\end{figure}

We can verify these predictions by performing some additional numerical simulations with TDF in the full Navier-Stokes equations. Figure \ref{fig:Q} (b) demonstrates that stability is observed for $\tau=5$ and $G^{\mathrm{max}}=0.1.$ Moreover $G^{\mathrm{max}}=0.5$ is less strongly stable, and $\tau=2$, $G^{\mathrm{max}}=0.1$ is more strongly unstable, all of which is consistent with the above analysis. While this stability analysis is useful in validating and optimising TDF, it does not offer much insight into, for instance, why $\tau\approx 5$ is optimal in this example. Following the ``frequency damping'' interpretation of TDF \citep{akervik_2006_steady,Shaabani2017}, we define the transfer function 

\begin{align*}
    H_{\mathrm{TDF}}(\mathrm{i}\omega,\tau) = \frac{1}{G} \frac{\mathcal{L}\{\bm{F}_{\mathrm{TDF}}\}}{\mathcal{L}\{\bm{u}\}} = e^{-\mathrm{i}\omega \tau}-1
\end{align*}
with $\mathcal{L}\{.\}(\omega) = \int_0^\infty . \,e^{-\omega t}dt$ the Laplace transform. 
The Laplace transform of the control term is $\mathcal{L}\{\bm{F}_{\mathrm{TDF}}\}(\omega) = -G(1-e^{-\omega \tau}) \mathcal{L}\{ \bm{u} \}.$ 
\modTY{Here,} $H_{\mathrm{TDF}}(\mathrm{i}\omega)$ allows us to ascertain the relative influence of the control term upon the temporal frequencies of the system in general (unlike a specific linear stability analysis). 
Figure \ref{fig:prodH} plots the magnitude of the transfer function against a normalised angular frequency. 
The first observation is that any zero mode is undamped by TDF control; this is an alternative explanation of the odd-number limitation where a purely real-valued eigenvalue cannot be stabilised in isolation. The next observation is that any harmonic of the feedback frequency $\frac{2\pi}{\tau}$ is also unaffected by the control. The consequence of these facts mean that any unstable eigenvalue of the target ECS with a frequency that is either close to zero, or close to the feedback frequency (or harmonic thereof) will not be stabilised. For high $Re$ travelling waves this means that careful tuning of the delay period, $\tau,$ may be necessary as the likelihood of an unstable eigenfrequency falling near a zero of the transfer function will increase.

We may use this analysis to interrogate the behaviour of TDF across a range of delay periods for the \textsf{UB} solution at $Re=2500$ discussed above. Defining $H_n(\tau) = |H_{\mathrm{TDF}}(\mathrm{i}\omega_n,\tau)|$ with $\omega_n$ the unstable eigenfrequencies of the uncontrolled solution as shown in figure \ref{fig:eigenvalues} (b), namely at $Re=2500,$ $\omega_1 = 0.75,\,\omega_2=0.48,\,\omega_3=0.32,$ we plot the product $H_1H_2H_3$ in figure \ref{fig:prodH}(b). 
This demonstrates clear agreement with the linear analysis of figure \ref{fig:TDF_LambdaMax} and therefore also the stabilisation of \textsf{UB} in figure \ref{fig:Q} (b). We see that the product of transfer functions is maximised at $\tau \approx 5$ and distinct zeros at harmonics of the eigenperiods, also coinciding with maxima of $\max(\mu)$ in figure \ref{fig:TDF_LambdaMax}. More specifically the first zero of $H_1H_2H_3$ is at $\tau=\frac{2\pi}{\omega_1} = 8.25$  in the same location as the first maximum of $\max(\mu)$ for these parameters, indicating, as expected, TDF will fail for poorly chosen time delays.

\subsection{Diagnosing unstable frequencies}
We have gained good insight into the behaviour of TDF in this example, however this analysis relies on knowledge of the unstable spectrum of the target solution. One goal \modDL{for a fully developed TDF method} would be the ability to obtain new, unknown solutions, without any knowledge of their properties in advance. In the example above we \modDL{began our investigation under this assumption}, 
i.e. \modDL{the values of} $\tau=2$ and $G^{\max}=0.5,$ used in figure \ref{fig:Q}(a), are set speculatively after a little trial and error. 

When $ |\lambda \tau | \ll 1,$ \modDL{we can} substitute $e^{-\lambda \tau} \approx 1-\lambda\tau$ in equation \eqref{eq:linear}, and we see that the effect of TDF on eigenvalues of $\mathcal{J}$ is a rescaling by $1+G\tau,$ \modDL{(i.e. the eigenvalue equation becomes $(1+G\tau)\lambda\bm{v}=\mathcal{J}\bm{v} $)}. \modDL{In the example above}, since $\tau$ is too small to effectively damp the lowest frequency eigenvalues, we observe \modDL{an unstable oscillation} with \modDL{modified} frequencies $\omega' \approx \frac{\omega}{1+G\tau}.$ In the \modDL{linear analysis of \textsf{UB}, instability} occurs at $Re\approx 2450$ with the $\tau=2$ and $G^{\max}=0.5$  TDF term active, with two complex conjugate pairs of eigenvalues crossing the imaginary axis with frequencies $\omega' \approx 0.16 \approx \omega_2/2$ and $\omega'\approx 0.246 \approx \omega_2/2.$ 
This means, if we can capture an estimate of these frequencies \modDL{in the numerical simulation where stabilisation has failed} it would be possible to \modDL{infer the original, unperturbed frequencies $\omega$}. This would enable one to perform the frequency domain analysis described earlier and choose more carefully tailored $\tau.$ 

Taking the time-series for the total kinetic energy $E(t)$ for the $Re=2500$ case with  $\tau=2$ and $G^{\max}=0.5,$ as shown in figure \ref{fig:Q}, and performing a Fourier transform, we observe, in figure \ref{fig:MTDF_TransferFunction} (b) a low-frequency/zero mode associated with a transient growth/decay of energy, but also two dominant modes at $0.246$ and $0.16,$ as predicted. Similarly we show the case with  $\tau=2$ and $G^{\max}=0.1,$ which has peaks in the power spectrum at $\omega_2/1.2\approx 0.4$ and $\omega_3/1.2\approx0.266,$ as we have predicted,  but also $\omega_2/1.2-\omega_3/1.2,$ indicating some nonlinear interactions taking place at slightly larger amplitude from the bifurcation. This approach \modDL{allows one to obtain} more effective time-delays when TDF fails to stabilise a solution and results in invasive, low dimensional behaviour, \modDL{near to a stabilisable state.} 

\section{Generalising time-delayed feedback control}
\label{sec:generalisingTDF}

\subsection{Extended time-delayed feedback}
In the analysis of the previous section we observe relatively narrow operating windows for which TDF will stabilise the target solution, which, as expected, decrease as $Re$ in increases. A common way to address this problem is via ``extended time-delayed feedback control'' (ETDF). This feeds an extended historical record into the control term by including times $t-k\tau$ for increasing integers $k$ as time evolves. For our implementation this would read, neglecting, for now, translations in $z,$ 
\begin{eqnarray}\label{eq:ETDF} 
F_{\mathrm{ETDF}}(t) 
&=& G \bigg[ (1-\epsilon)\sum_{k=1}^{\infty} \epsilon^{k-1} u(t-k\tau) - u(t) \bigg] \:, \notag \\
&=& G [ u(t-\tau) - u(t) ] + \epsilon F_{\mathrm{ETDF}}(t-\tau) \:, 
\end{eqnarray}
where $0 \le \epsilon \le 1$, see \citet{Socolar1994}. When $\epsilon=0$ the control reduces back to the standard Pyragas control. This has the effect of broadening and flattening the transfer function:
\begin{align*}
    H_{\mathrm{ETDF}}(\mathrm{i}\omega,\tau) = \frac{1}{G} \frac{\mathcal{L}\{F_{\mathrm{ETDF}}\}}{\mathcal{L}\{u\}} = \frac{e^{-\mathrm{i}\omega \tau}-1}{1-\epsilon e^{-\mathrm{i}\omega \tau}},
\end{align*}
see figure \ref{fig:prodH}. This method is attractive as it is relatively simple to implement and in practice only requires storing one additional history array for $F_{\mathrm{ETDF}}(t-\tau)$ (over a period). However, for our case, in particular for the analysis performed above with \textsf{UB}, we observe that the flattening of the transfer function results in much less distinct maxima, in particular the peak at $\tau=5.5$ is much smaller (Figure \ref{fig:prodH}) and is unlikely to result in the same stabilisation observed with regular TDF. Moreover we have seen that attempting to stabilise states with several unstable eigenfrequencies is challenging, and ETDF gives only a marginal advantage in this regard, particularly if the frequencies are well separated. 


\subsection{Multiple time-delayed feedback} \label{sec:MTDF}
In the previous example for \textsf{UB} the instability observed at $Re=2500$ using $\tau=2$ was remedied following a careful analysis of the effect of TDF terms on the unstable eigenvalues of the uncontrolled system. It was noted that the frequencies are key in choosing an appropriate time-delay and that they may be observed in unsuccessful, invasive TDF dynamics. 

\begin{figure}
\centering
\subfigure{
\begin{minipage}{0.5\linewidth}
\centerline{(a)}
\includegraphics[width=\linewidth]{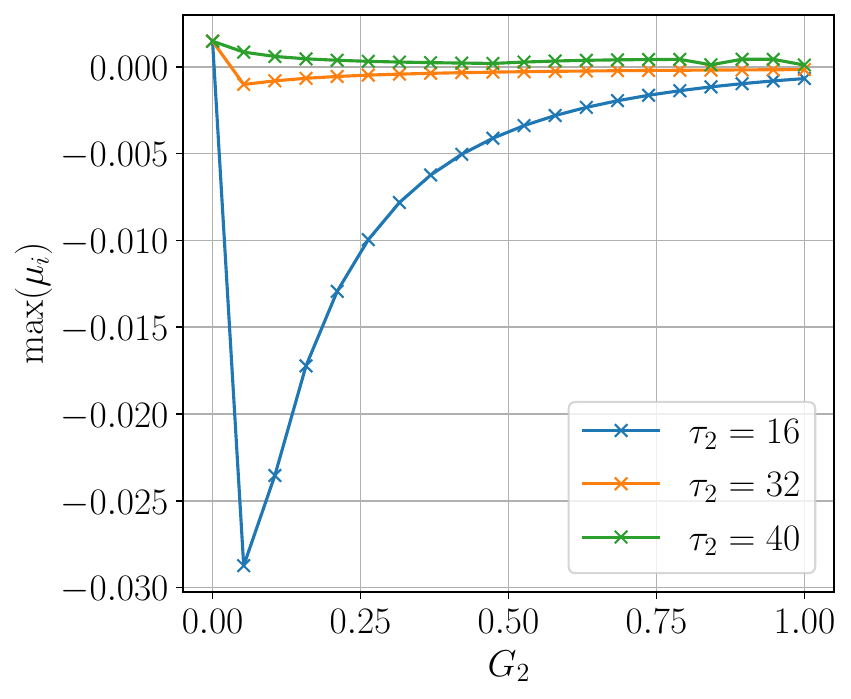}
\end{minipage}
\begin{minipage}{0.5\linewidth}
\centerline{(b)}
\includegraphics[width=\linewidth]{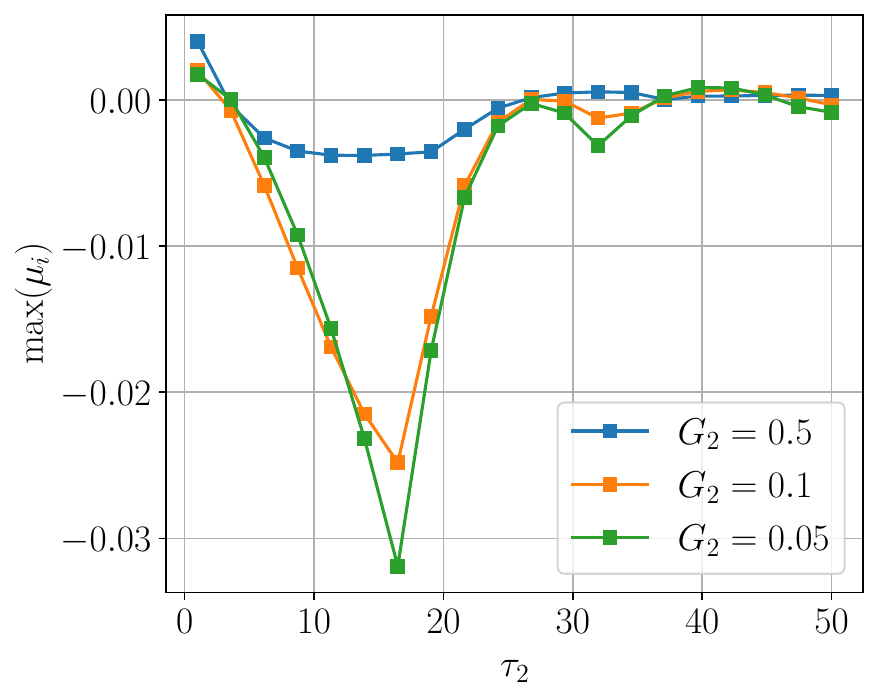}
\end{minipage}
}
\caption{Dependence of the largest real-part of the eigenvalue spectrum $\max_i \mu_i$ with $G_2$ and $\tau_2,$ for $Re=2500$ \textsf{UB} solution with two-term MTDF and $\tau_1=2$ and $G_1=0.5$. Most effective stabilisation is observed with $\tau_2\approx 16$ and the rather modest $G_2\approx 0.05.$
}
\label{fig:MTDF_LambdaMax}
\end{figure}

\begin{figure}
\centering
\begin{minipage}{0.39\linewidth}
\centerline{(a)}
\includegraphics[width=\linewidth]{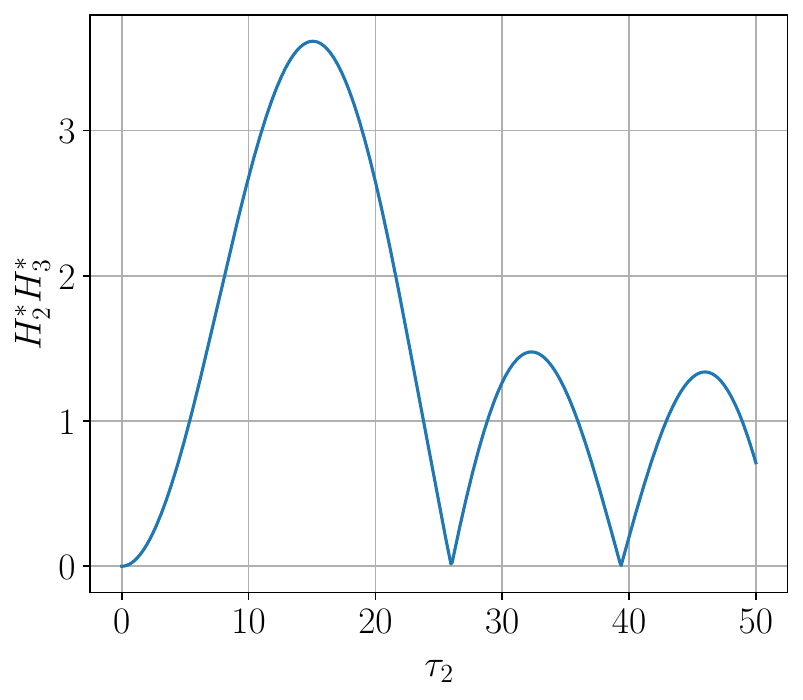}
\end{minipage}
\begin{minipage}{0.6\linewidth}
\centerline{(b)}
\includegraphics[width=\linewidth]{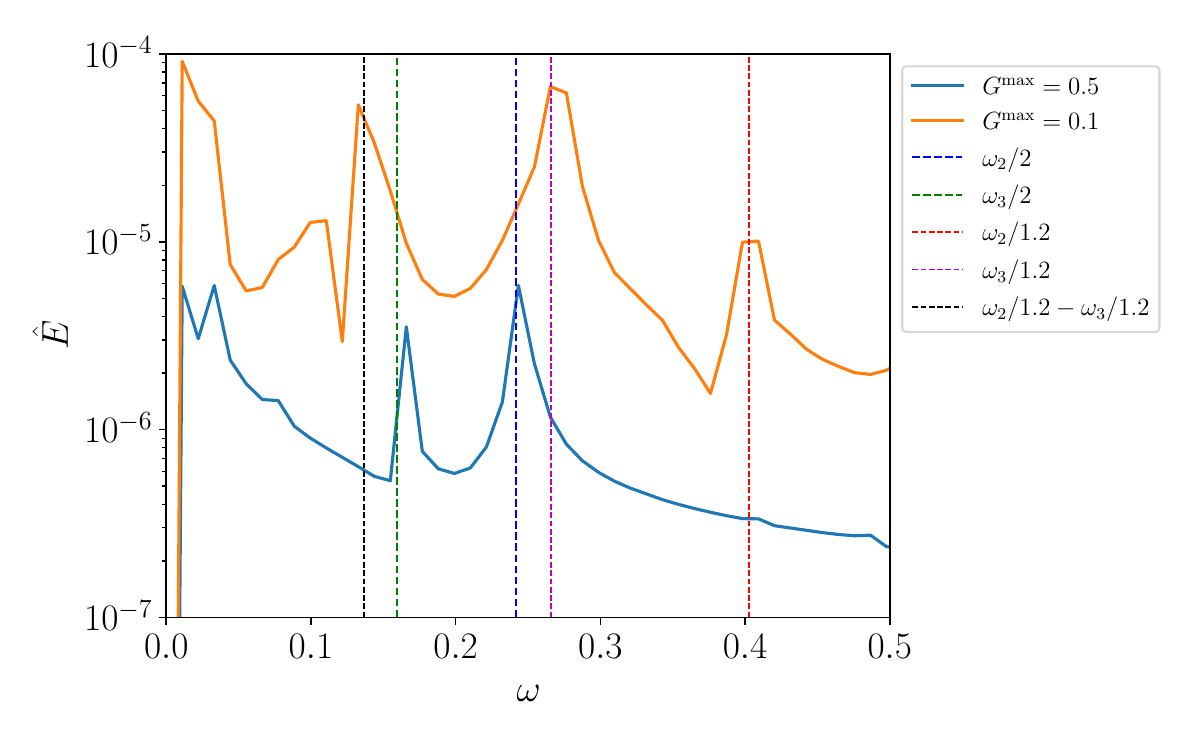}
\end{minipage}
\caption{ (a) shows the product of transfer functions $H_1^*H_2^*$ as defined in the text, for the modified eigenfrequencies when attempting to stabilise \textsf{UB} at $Re=2500$ using $\tau=2$ and $G=0.5.$ Note that the peak coincides with figure \ref{fig:MTDF_LambdaMax}b, i.e. $\tau_2\approx 16$ is around the optimal. 
(b) shows the power spectrum of energy, $\hat{E}$, for the unsuccessful TDF cases at $Re=2500$ with $\tau=2$ and $G = 0.5$ and 0.1. Vertical lines show that the peaks in these spectra follow the expected scaling by $1/(1+\tau G).$
}
\label{fig:MTDF_TransferFunction}
\end{figure}

\begin{figure}
\centering
\subfigure{
\begin{minipage}{0.5\linewidth}
\centerline{(a)}
\includegraphics[width=\linewidth,clip]{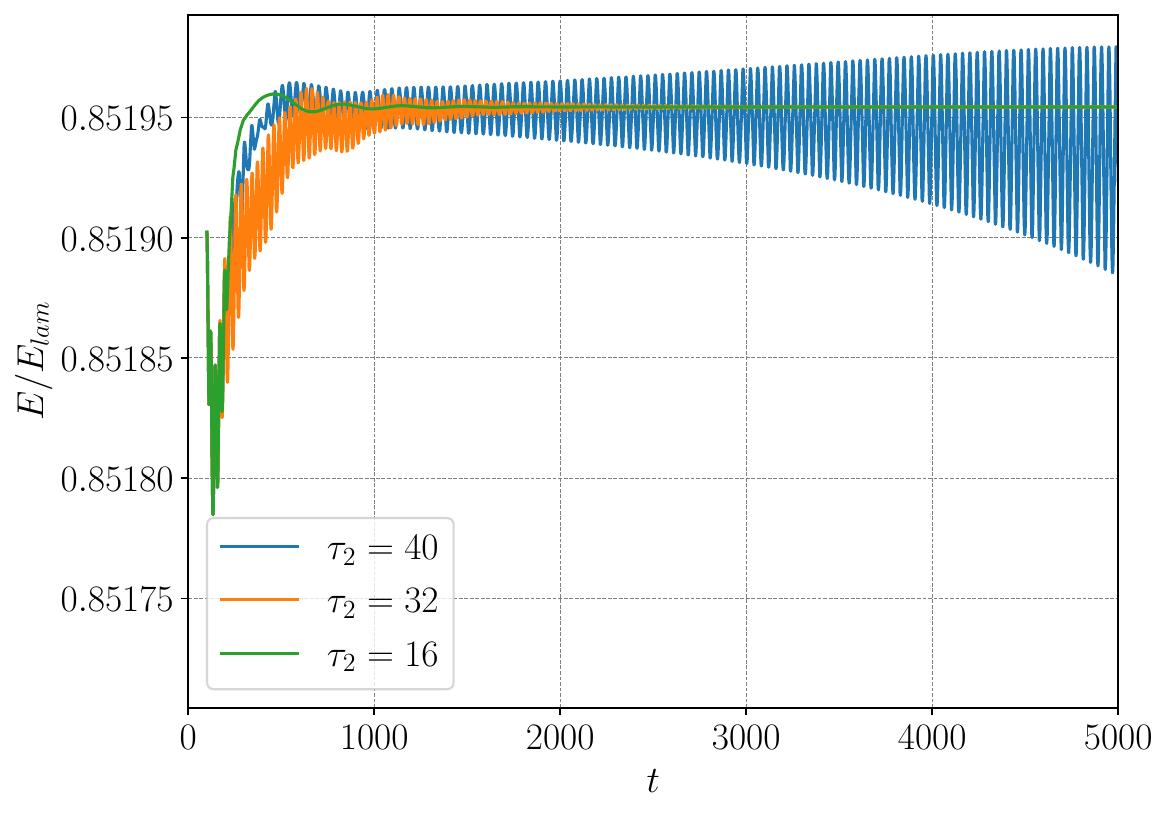}
\end{minipage}
\begin{minipage}{0.5\linewidth}
\centerline{(b)}
\includegraphics[width=\linewidth,clip]{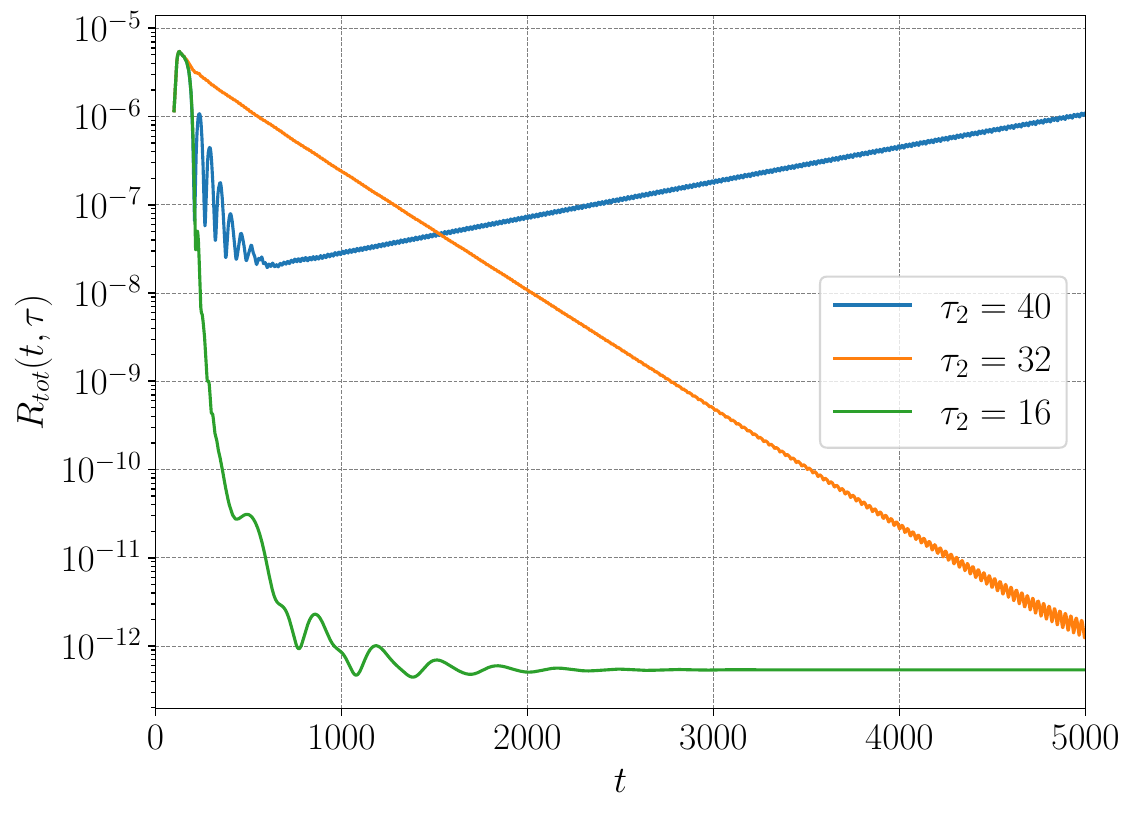}
\end{minipage}
}
\caption{
MTDF stabilisation of \textsf{UB} travelling wave at $Re=2500$, 
shown via time series of $E/E_{\mathrm{lam}}$ (a) and $R_{\mathrm{tot}}$ (b). 
Comparison shows various choices of $\tau_2$ with $\tau_1=2,$ $G_1=0.5,$ and $G_2=0.1.$ We see rapid stabilisation for the optimal $\tau_2=16,$ moderate stabilisation for $\tau_2=32$ and instability for $\tau=40.$
}
\label{fig:MTDF}
\end{figure}

To attenuate oscillations across several frequencies, 
we introduce multiple time-delayed feedback (MTDF), 
$\bm{F}_{\mathrm{MTDF}}$, such that
\begin{eqnarray}\label{eq:MTDF}
\bm{F}_{\mathrm{MTDF}}(r,\theta,z,t) 
= \sum_i^N G_i(t) 
[\mathcal{T}_{z}(s_z) \bm{u}(r,\theta,z,t-\tau_i)-\bm{u}(r,\theta,z,t)] \:,  
\end{eqnarray}
where the $i$th time delay is $\tau_i$ ($i=1,2,\cdots,N$) and 
we translate the $i$-th delayed state, $\bm{u}(r,\theta,z,t-\tau_i)$, 
by the streamwise shift, $s_z(t;\tau_i)$. 
\modTY{We adjust each streamwise shift for every delay term using the gradient descent method outlined in \S~\ref{ss:formulation}, } \modDL{however we have verified that adapting the $i=1$ term only and using $s_z(t;\tau_i) = c_z(t)\tau_i$, to match the phases speeds of the other terms is as effective.}
%
Delayed feedback with multiple delays 
was previously used in the low-dimensional chaotic system 
of Chua's circuit by \citet{ahlborn_2004_stabilizing}, 
to stabilise a nonlinear steady solution.
They reported that multiple time-delayed feedback helped 
to expand the basin of attraction of equilibrium solutions, 
being more efficient than extended time-delayed feedback (ETDF) 
\citep{Socolar1994,sukow_1997_controlling}. 
We can repeat the analysis of \S~\ref{sec:linear} with MTDF by simply introducing a second TDF term into equation \eqref{eq:linear}. We will begin by investigating the effect of introducing a second term to the unstable $Re=2500$ case, i.e. $\tau_1=2,$ $G_1=0.5$ and varying $\tau_2$ and $G_2.$
Given that we have observed $\omega_2/2$ and $\omega_3/2$ in the TDF example, it would be sensible to choose a $\tau_2$ to stabilise these modes. Via the transfer function analysis of the previous section, now with $H_n^* = |H_{\mathrm{TDF}}(\mathrm{i}\omega_n/(1+G_1\tau_1),\tau_2)|,$ we are again able to predict an optimal $\tau_2$. 
The maxima of the product $H_2^*H_3^*$ line up 
with the minima of $\max(\mu_i)$ 
from the MTDF linear theory, indicating an optimal $\tau_2\approx 16$, 
see figures~\ref{fig:MTDF_LambdaMax}b and \ref{fig:MTDF_TransferFunction}a. 
The interpretation here is that the second term should act to attenuate the frequencies modified by the first term (or vice-versa). It should also be noted that there is now a far larger range of $\tau_2$ and $G_2$ which stabilises \textsf{UB}, meaning less speculative tuning of parameters, particularly of $G_2.$
\modDL{This result is confirmed in the numerical simulation, figure \ref{fig:MTDF} shows very fast stabilisation of \textsf{UB} at $Re=2500$ with $\tau_1=2$ and $\tau_2=16.$ Success is also observed with $\tau_2=32$ although with a slower rate of attraction, 
and $\tau_2=40$ shows instability, all of which is consistent with the linear and frequency analysis shown in figures \ref{fig:MTDF_LambdaMax} and \ref{fig:MTDF_TransferFunction}.}

Note one could define a transfer function $H_{\mathrm{MTDF}},$ for MTDF, given by
\begin{eqnarray}\label{eq:MTDF_TF} 
H_{\mathrm{MTDF}}(\mathrm{i}\omega) 
= \frac{1}{\sum_i^N G_i} \frac{\mathcal{L} \{F_{\mathrm{MTDF}}\}}{\mathcal{L} \{u\}} 
= \frac{\sum_i^N G_i\exp(-i\omega\tau_i)}{\sum_i^N G_i}-1 
\end{eqnarray}
However applying this to the case under discussion effectively yields a repeat of the single TDF result, in that the optimal $\tau_2$ to stabilise the three eigenfrequencies, with $\tau_1=2,$ is around 5; $H_{\mathrm{MTDF}}$ does not accurately predict the effect of combinations of time-delays upon a given state.


\section{Stabilisation of nonlinear travelling wave from turbulence} 
\label{sec:TDF_turb}
%
Having described the ability of MTDF to successfully damp instability of nonlinear travelling waves, in this section we will demonstrate that such stabilised states can have suitably large basins of attractions and that it is possible to stabilise them from turbulence.
We proceed as if the properties of the target solutions are unknown and allow the multiscale fluctuations of turbulence to fully develop before attempting stabilisation using MTDF. Turbulent fluctuations may include much longer timescales than the eigenperiods of the target solution. We imagine that some well chosen delays can suppress sufficient spatiotemporal fluctuations in turbulence, without relaminarising the flow completely, at the same time as achieving stabilisation of the target solution. 
In other words, there is some motivation for including MTDF terms to widen the basin of attraction of the target state, as well as to force eigenvalues across the imaginary axis.
However, this introduces a challenge: the more delays we use, the more parameters we need to adjust to stabilise nonlinear travelling waves successfully. 

Without prior knowledge of the solution's instability, we seek to exploit some automatic techniques for obtaining stabilising parameter values. 
We have seen how $\tau_i$ may be chosen by analysing the data; in the next section we discuss the adaptive gain method of \modTY{\citet{Lehnert:2011hu}}, 
which automates the selection of $G$ and seeks to avoid a laborious parameter search.
%

\subsection{Adaptive gain method}
\label{SS:adaptG}
The speed-gradient method of \cite{Lehnert:2011hu} seeks to find an optimal TDF gain by dynamically adjusting $G(t)$ by a ``speed-gradient'' descent method. 
In order to exploit this method, we need to extend it to handle multiple terms and the translation operator which is applied to the control term(s). We define the cost function, $Q_i(t)$, for the $i$-th feedback term as: 
\begin{eqnarray}\label{eq:AGM3} 
Q_i(t) = \frac{1}{2} \int |
\mathcal{T}_z(s_z) \bm{u}(\bm{x},t-\tau_i)- \bm{u}(\bm{x},t)|^2 \mathrm{d}V \:, 
\end{eqnarray}
where $\bm{x}$ is the position vector and 
successful control yields $Q_i(t) \rightarrow 0$ as $t \rightarrow \infty$. 
The speed gradient algorithm 
in the differential form is given by 
\begin{eqnarray}\label{eq:AGM4} 
\frac{\mathrm{d}}{\mathrm{d} t}{G}_i(t) = - \gamma_i^G \nabla_{G_i} \frac{\mathrm{d}{Q}_i}{\mathrm{d} t} \:, 
\end{eqnarray}
where $\gamma_i^G>0$ is a free parameter controlling the descent rate 
and $\nabla_{G_i}$ denotes $\partial/\partial G_i$. 
By taking time-derivative of (\ref{eq:AGM3}), we have 
\begin{eqnarray}\label{eq:AGM5} 
\frac{\mathrm{d}{Q}_i}{\mathrm{d} t} =  \int g_i(\bm{x},t) \mathrm{d}V \:, 
\end{eqnarray}
where
\begin{multline}\label{eq:AGM6}
g_i(\bm{x},t) = 
[
\mathcal{T}_z(s_z) 
\bm{u}(\bm{x},t-\tau_i)- \bm{u}(\bm{x},t)]\cdot
\bigg[
\mathcal{T}_z(s_z) 
\frac{\partial \bm{u}}{\partial t}(\bm{x},t-\tau_i)
-\frac{\partial \bm{u}}{\partial t}(\bm{x},t)\bigg] 
\end{multline}
and we have assumed that $\gamma_s$ is chosen such that $\frac{\mathrm{d}s_z}{\mathrm{d} t} \ll 1,$ hence $s_z$ is approximately constant for the purposes of updating $G_i(t).$
Using the Navier--Stokes momentum equations with MTDF in the form
\begin{align}\label{eq:AGM1} 
\frac{\partial}{\partial t} \bm{u}(\bm{x},t) = 
\bm{f}(\bm{x},t) + \sum_i^N G_i [
\mathcal{T}_z(s_z) \bm{u}(\bm{x},t-\tau_i) - 
\bm{u}(\bm{x},t)] \:, 
\end{align}
where $\bm{f}$ includes all the terms from the right-hand side of the momentum equation, alongside 
%
(\ref{eq:AGM4}) and (\ref{eq:AGM5}), 
we obtain the following formula: 
\begin{eqnarray}\label{eq:AGM7} 
\frac{\mathrm{d}}{\mathrm{d} t}{G}_i(t) = - \gamma_i^G \int h_i(\bm{x},t) \mathrm{d} V, 
\end{eqnarray}
where 
\begin{multline}
\label{eq:AGM8} 
h_i(\bm{x},t) 
= [
\mathcal{T}_z(s_z) 
\bm{u}(\bm{x},t-\tau_i)- \bm{u}(\bm{x},t)] \cdot \\
[\bm{u}(\bm{x},t) - 2 
\mathcal{T}_z(s_z) \bm{u}(\bm{x},t-\tau_i) 
+ 
\mathcal{T}_z(2s_z) \bm{u}(\bm{x},t-2\tau_i)] \:.  
\end{multline}
Note that on taking $\nabla_{G_i}$ only the MTDF terms contribute to $h_i.$
Equation (\ref{eq:AGM8}) indicates that 
the code should store instantaneous velocity field data over 2 delay periods, $2 \tau_i,$ when using this adaptive gain method, hence doubling the storage requirements. 
However, we implement a temporal interpolation procedure using cubic splines \citep{Shaabani2017} 
 to enable us to store longer historical records of $\bm{u},$ albeit at the cost of some approximation error in the interpolation method.
%

\subsection{\textsf{UB} with MTDF and adaptive gain.}\label{SS:TDFresults}
%
Here we attempt to stabilise \textsf{UB} at $Re=3000$ 
from a turbulent state,
in the ($S$,$Z_2$)-symmetric subspace using MTDF. 
In this subspace, 
\textsf{UB} at $Re=3000$ has 4 pairs of complex unstable eigenvalues, and 
therefore it is unaffected by the odd-number limitation \citep{Nakajima1997}. 
We approach this problem as if the eigenvalues are unknown and we select $\tau_i$s and other MTDF parameters without any linear analysis to steer our choices.

\begin{figure}
\centering
\subfigure{
\begin{minipage}{0.5\linewidth}
\centerline{(a)}
\includegraphics[width=\linewidth,clip]{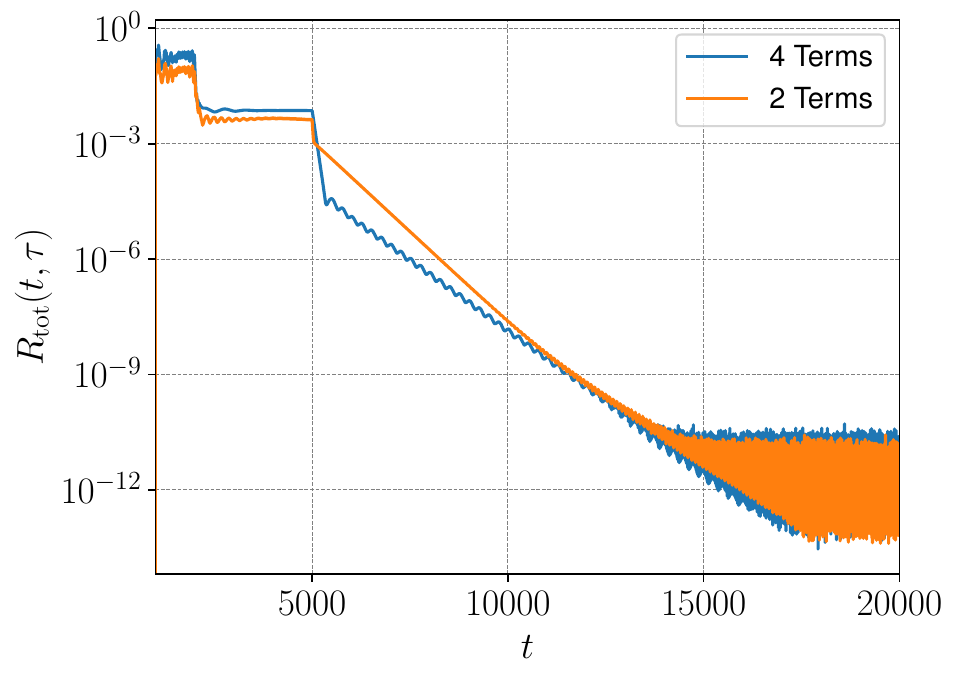}
\end{minipage}
\begin{minipage}{0.5\linewidth}
\centerline{(b)}
\includegraphics[width=\linewidth,clip]{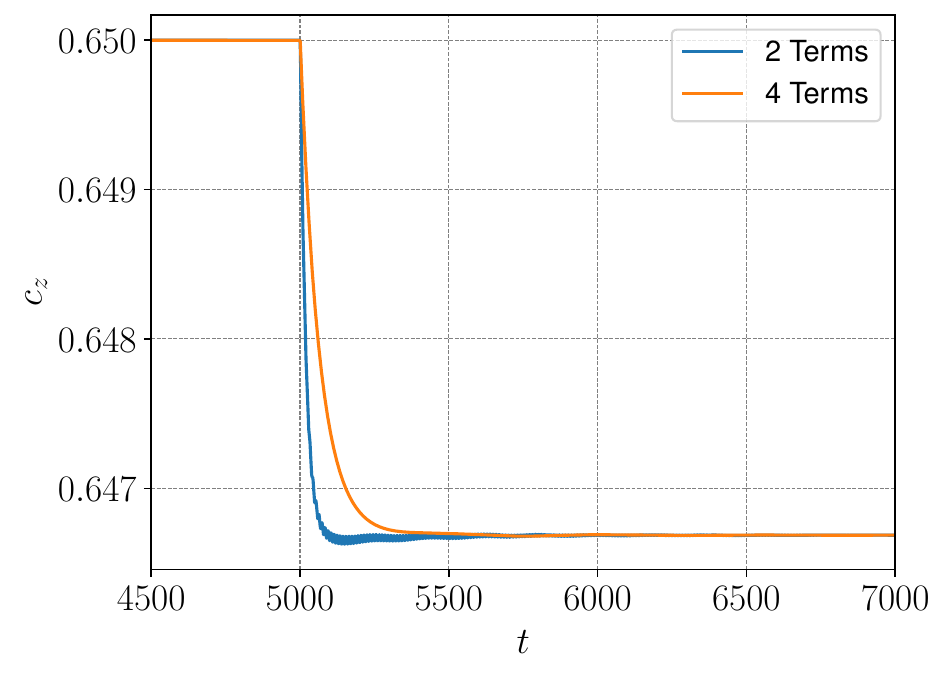}
\end{minipage}}
\subfigure{
\begin{minipage}{0.49\linewidth}
\centerline{(c)}
\includegraphics[width=\linewidth,clip]{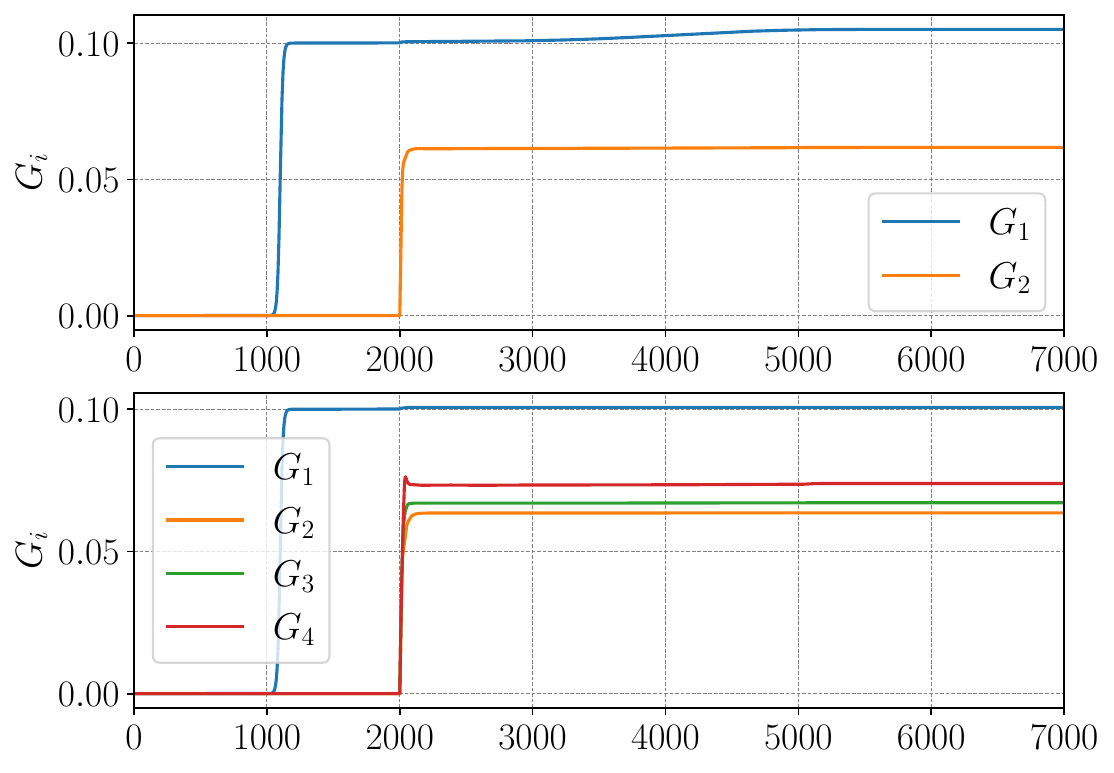}
\end{minipage}
\begin{minipage}{0.51\linewidth}
\centerline{(d)}
\includegraphics[width=\linewidth,clip]{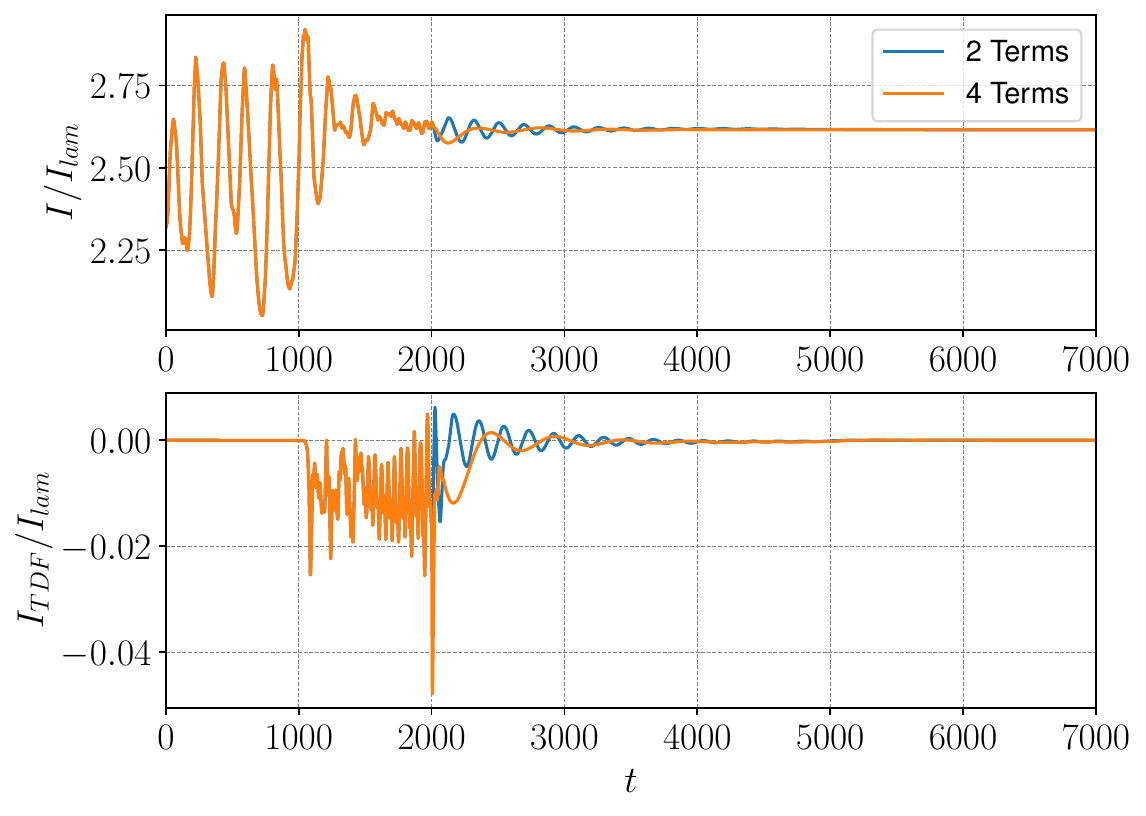}
\end{minipage}
}
\caption{
Successful stabilisation of \textsf{UB} at $Re=3000$ 
from a turbulent state using MTDF (\ref{eq:MTDF}) 
with two terms ($\tau_1=2$, $\tau_2=9$) and four terms ($\tau_1=2$, $\tau_2=9$, $\tau_3=17,$ $\tau_4=33$). 
(a) $R_{\mathrm{tot}}$, 
(b) $c_z$, 
\modTY{
(c,top) $G_i$ with two terms,
(c,bottom) $G_i$ with four terms, 
(d) $I/I_{\mathrm{lam}}$ and $I_{\mathrm{TDF}}/I_{\mathrm{lam}}$. 
}
The gain, $G_1(t)$, is switched on at $t=1000$. 
Here, $G_1(t)$ is initially increased following the sigmoid gain function (\ref{eq:sigmoidgain}) where ($t_s$,$G_1^{\max}$,$a$,$b$)=(1000,0.1,0.1,100). 
At $t=2000$, the $G_i$ parameters begin to evolve using the adaptive gain method as described by equation \eqref{eq:AGM8}, with a constant value of $\gamma_i^G=0.1$ applied to all delay terms.
The adaptive shift method is switched on at $t=5000$, 
where $\gamma^s=0.1$ and $c_z(0)=0.65$. 
}
\label{fig:MTDF_turb_Re3000}
\end{figure}

\begin{figure}
\centering
\subfigure{
\begin{minipage}{0.48\linewidth}
\centerline{(a)}
\includegraphics[width=\linewidth,clip]{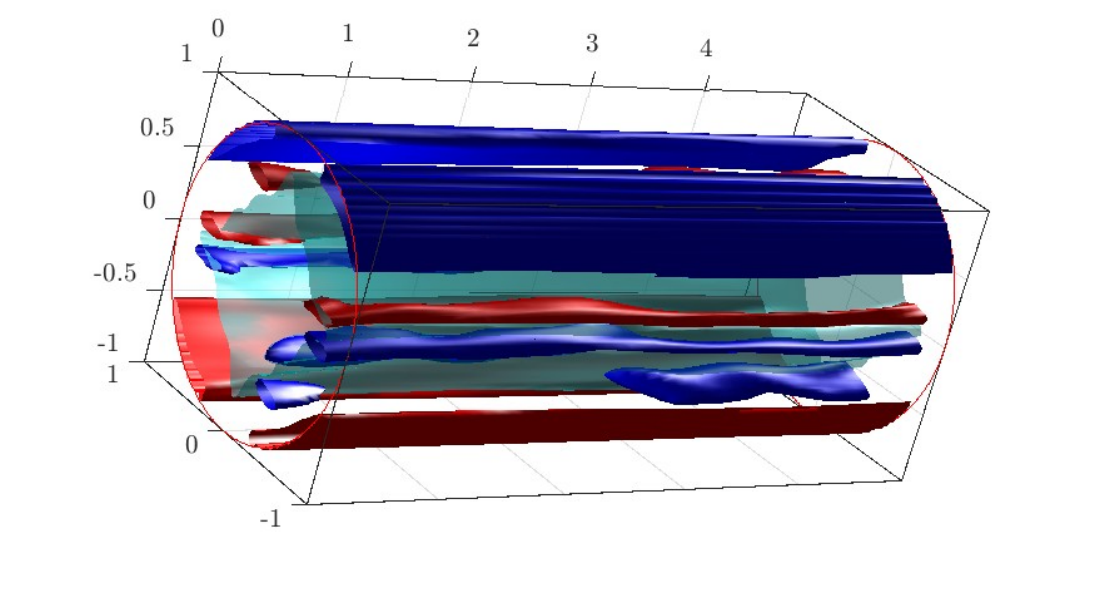}
\end{minipage}
}
\subfigure{
\begin{minipage}{0.48\linewidth}
\centerline{(b)}
\includegraphics[width=\linewidth,clip]{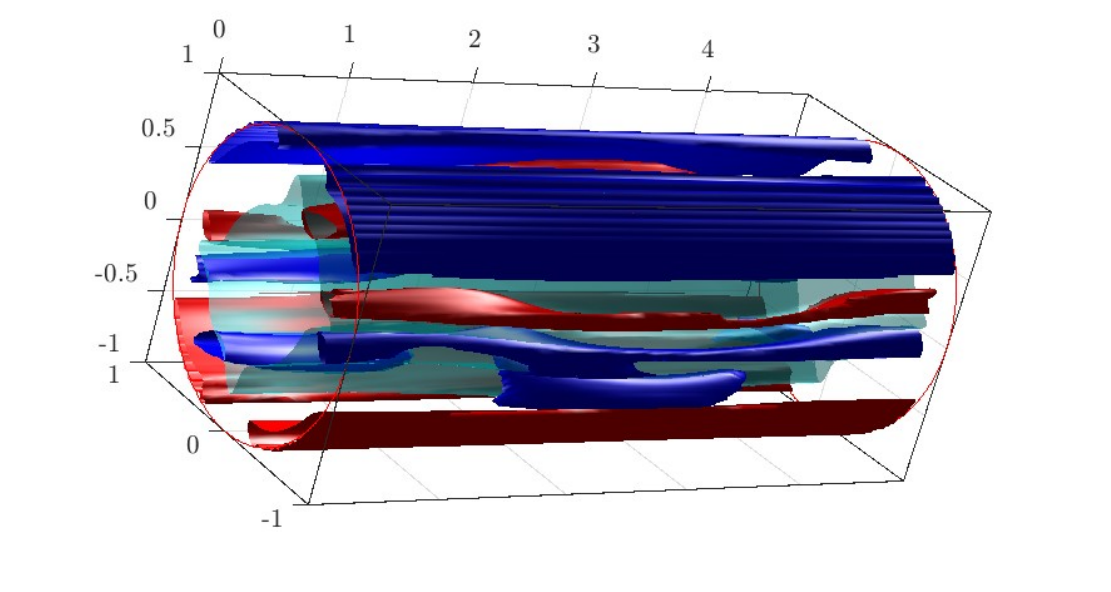}
\end{minipage}
}
\caption{
Two snapshots from the simulation using MTDF with two delays 
(see figure~\ref{fig:MTDF_turb_Re3000}). 
(a) Turbulent field at $t=0$ and 
(b) stabilised \textsf{UB} at $t=20,000$.  
The cyan isosurface denotes $u_z=-0.1$. 
The red and blue isosurfaces denote 
$\omega_z=0.15$ and $-0.15$, respectively, 
where $\omega_z = (\nabla\times\bm{u})_z$ is the streamwise fluctuating vorticity. 
}
\label{fig:mtdfc_snapshots}
\end{figure}

We consider first an MTDF case with two delays, $(\tau_1,\tau_2)=(2,9)$. 
We set the following parameters for $\tau_1$; ($t_s$,$G^{\max}$,$a$,$b$,$\gamma^s$)=(1000,0.5,0.1,100,0.1), i.e. TDF becomes active at $t=1000,$ giving a period of turbulent activity, and an initial sigmoid profile of $G(t)$. The adaptive gain method for $G_1(t)$, with $\gamma_1^G=0.1$, is then started at $t=2000$, and $G_1(t)$ evolves following the ODE (\ref{eq:AGM4}). The second feedback term with gain, $G_2(t)$, evolves from zero at $t=2000$ through the adaptive gain method ($\gamma_2^G=0.1$). 
Without knowing successful values of $G_i$ and $s_z(t;\tau_i)$ in advance, we successfully stabilise \textsf{UB} at $Re=3000$ using this double-delay MTDF (see figure~\ref{fig:MTDF_turb_Re3000}). 
In order to verify the non-invasive nature of the stabilisation in the MTDF cases, 
the definition of $R_{\mathrm{tot}}$ requires updating
\begin{equation} \label{eq:residual}
R_{\mathrm{tot}}
= \frac{\|\sum_i^N\left[\mathcal{T}_z(s_z)\bm{U}(r,\theta,z,t-\tau_i)-\bm{U}(r,\theta,z,t)\right]\|_2}
{\|\bm{U}(r,\theta,z,t)\|_2}.
\end{equation}
Successful, non-invasive stabilisation 
is quantitatively confirmed by the very small values 
of both $R_{\mathrm{tot}}$ (figure \ref{fig:MTDF_turb_Re3000} (a)) and $I_{\mathrm{TDF}}/I_{\mathrm{lam}}$,  
the latter being of order of $10^{-8}$ at $t=10000$ 
(see figure~\ref{fig:MTDF_turb_Re3000} (d)). 
\modDL{The phase speed} $c_z$ is also evolved onto its exact value 
through the adaptive translation method and we observe $G_i$ 
being adjusted by the speed-gradient method onto stabilising values. 
A snapshot of the stabilised \textsf{UB} at $t=20000$ 
is shown in figure~\ref{fig:mtdfc_snapshots}(b) 
with a turbulent field at $t=0$ for comparison in figure~\ref{fig:mtdfc_snapshots}(a). 
%

Figure~\ref{fig:MTDF_turb_Re3000} also shows an MTDF case 
with four delays where $\tau_1=2$, $\tau_2=9$, $\tau_3=17$, $\tau_4=33$.
%
Motivated by our frequency-domain analysis in \S~\ref{sec:linear}, these delays are carefully chosen to be non-commensurate and provide broad coverage of the temporal spectrum.
%
%
As in the two-delay case, 
only $G_1(t)$ is switched on at $t=1000$ 
following the sigmoid gain function (\ref{eq:sigmoidgain}) , 
where ($t_s$,$G_1^{\max}$,$a$,$b$,$\gamma^s$)=(1000,0.1,0.1,100,0.1). 
The adaptive gain method is switched on at $t=2000$ with $\gamma^G=0.1$ for all the terms, . 
Once the time-dependent gains are turned on, 
$I_{\mathrm{TDF}}/I_{\mathrm{lam}}$ starts to 
fluctuate with the amplitude smaller than 0.02 and 
then exhibits a damped oscillatory behaviour, 
similar to the case with two delays. There are no significant qualitative differences 
between \modTY{the two- and four-term MTDF cases}; 
in the four delay case $G_3$ is the second largest (recall $\tau_3=17$) and $G_2$ takes a smaller value at stabilisation than the two-delay case. \modDL{Gain} $G_4$ is the smallest but still make a significant contribution.  This suggests that the terms are all influencing the stabilisation, while the two-delay case demonstrates that not all terms are necessary for stabilisation. \modDL{Figure~\ref{fig:MTDF_turb_Re3000} (a) shows that} the rate of attraction is of the same order in both cases, \modDL{saturation is reached at roughly the same time,} this may suggest that the principle benefit of additional terms in MTDF is to widen the parameter windows under which \textsf{UB} is stabilised.
If stabilising $G$ are not obtained, it may be observed that the state is brought near the target solution, but then moves away from it along some unstable manifold, similar to the blue curves in figure~\ref{fig:MTDF}. The adaptive method gives the gains freedom to find values which succeed in stabilisation, instead of this close approach.
%
It is worth noting that the energy injected by the forcing is only around 2-3\% of the energy injected by the imposed pressure gradient, see figure \ref{fig:MTDF_turb_Re3000}(d). This means that, even at early times when the MTDF terms are largest, the overall energetic influence of MTDF on the flow is quite small. This may provide motivation for development of TDF or MTDF in real-world experimental control situations; TDF does not need to intervene strongly to stabilise these travelling waves.

\begin{figure}
\centering
\subfigure{
\begin{minipage}{0.49\linewidth}
\centerline{(a)}
\includegraphics[width=\linewidth,clip]{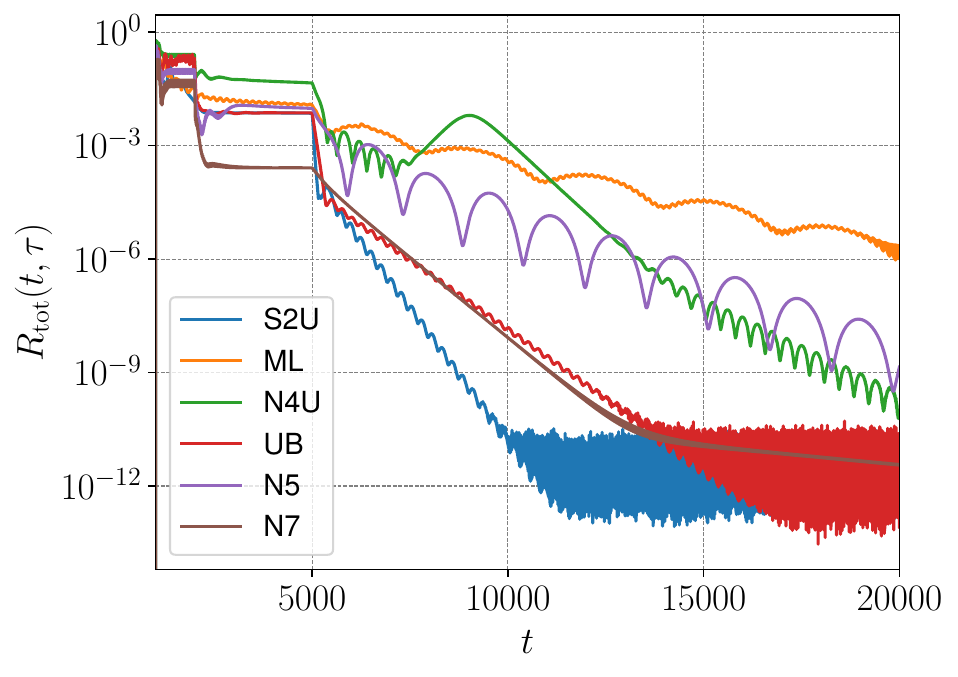}
\end{minipage}
\begin{minipage}{0.52\linewidth}
\centerline{(b)}
\includegraphics[width=\linewidth,clip]{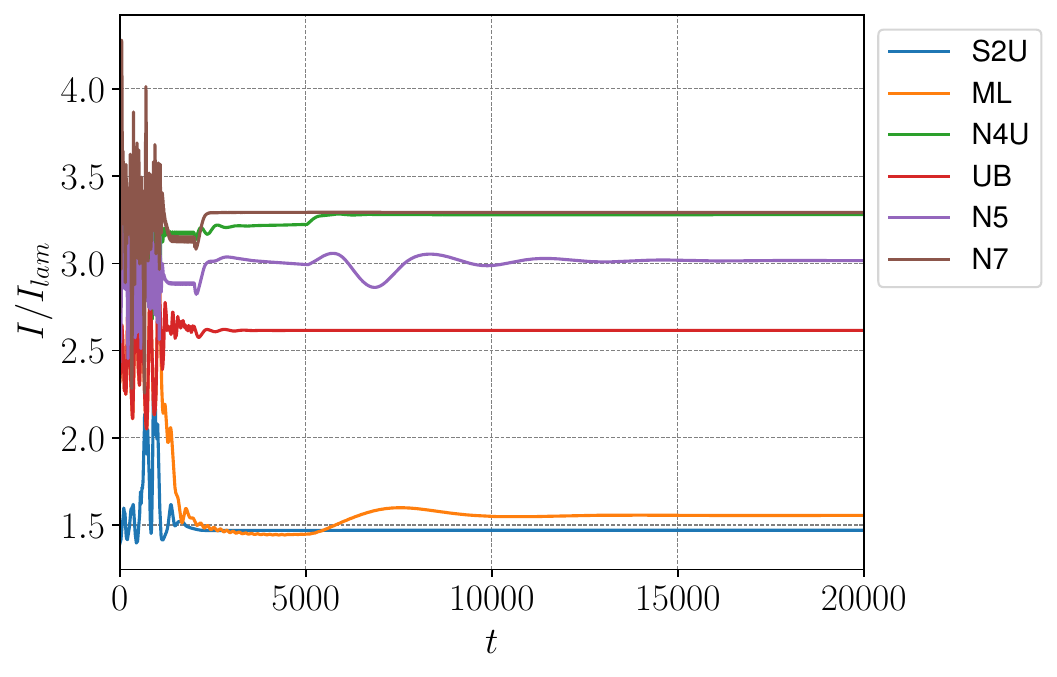}
\end{minipage}}
\subfigure{
\begin{minipage}{0.49\linewidth}
\centerline{(c)}
\includegraphics[width=\linewidth,clip]{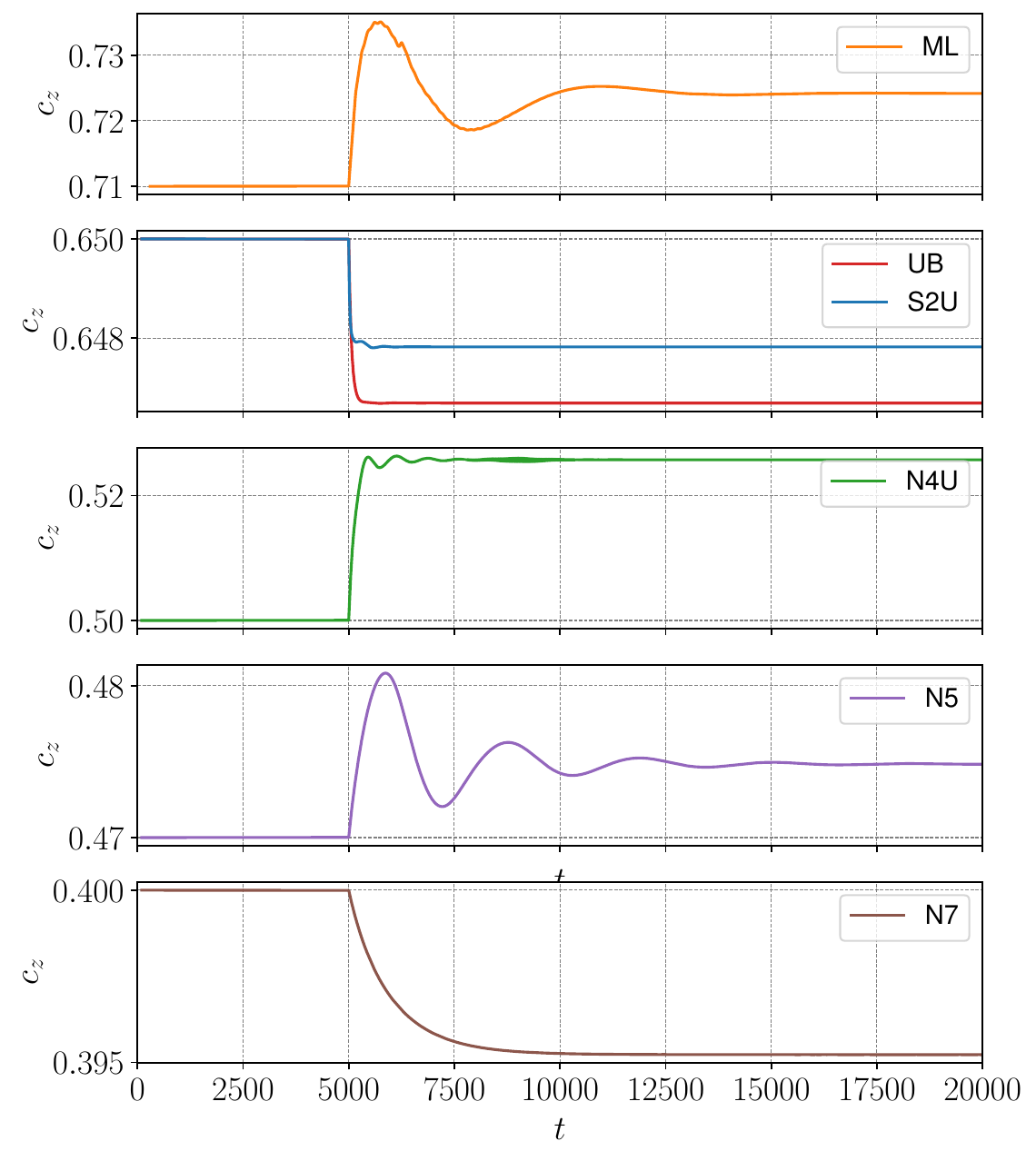}
\end{minipage}
\begin{minipage}{0.52\linewidth}
\centerline{(d)}
\includegraphics[width=\linewidth,clip]{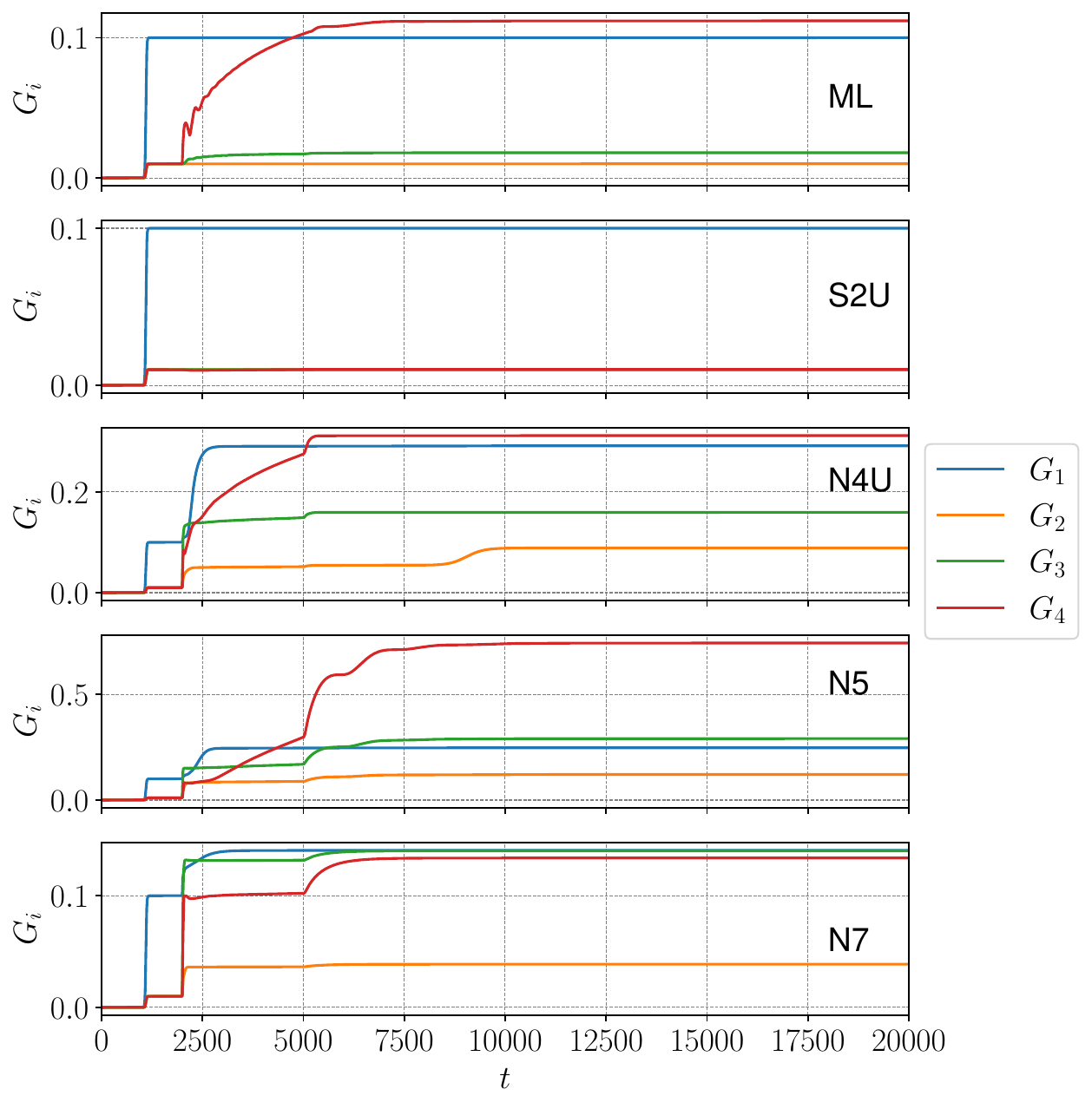}
\end{minipage}
}
\caption{
Plots showing successful stabilisation of \textsf{S2U, ML, N4U, N5, N7} in their respective subspaces, from a turbulent state using four-term MTDF (\ref{eq:MTDF}). Specific parameter values can be found in table \ref{tab:2}. 
(a) Shows the residual $R_{\mathrm{tot}},$ (b) the energy input $I/I_{\mathrm{lam}},$ (c) the phase speeds $c_z$ dynamically converging to their exact values, and (d) the time-dependent gains $G_i(t).$
}
\label{fig:MTDF_turb_all}
\end{figure}
\begin{figure}
\centering
\centerline{(a)}
\includegraphics[width=\linewidth,clip]{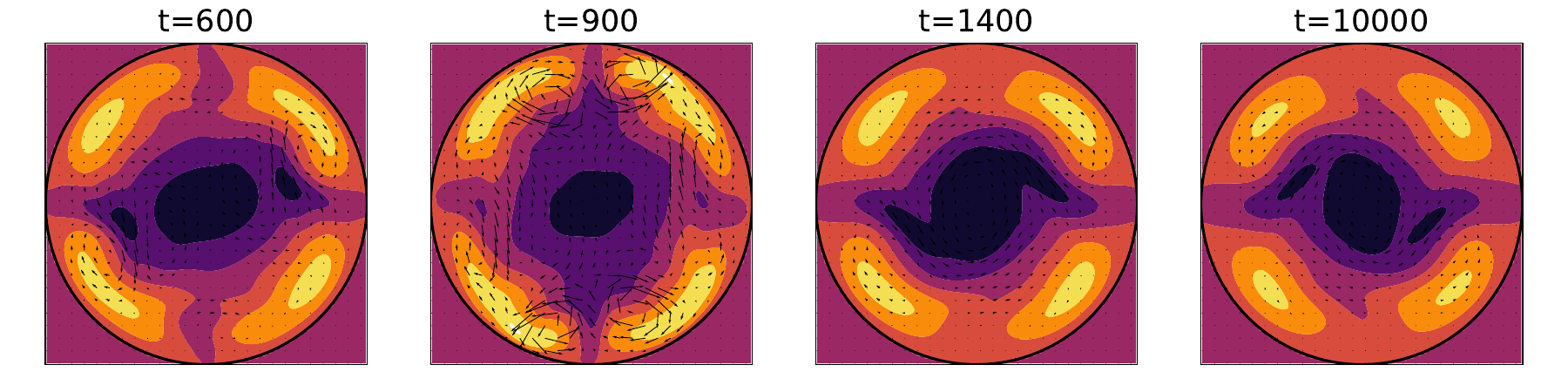}\\
 \centerline{(b)}
\includegraphics[width=\linewidth,clip]{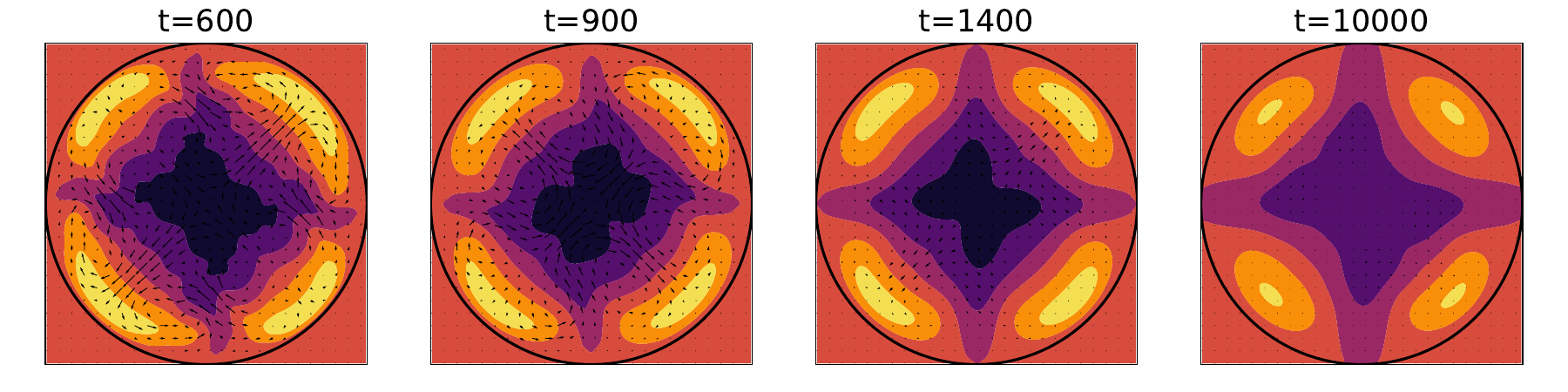}\\
\centerline{(c)}
\includegraphics[width=\linewidth,clip]{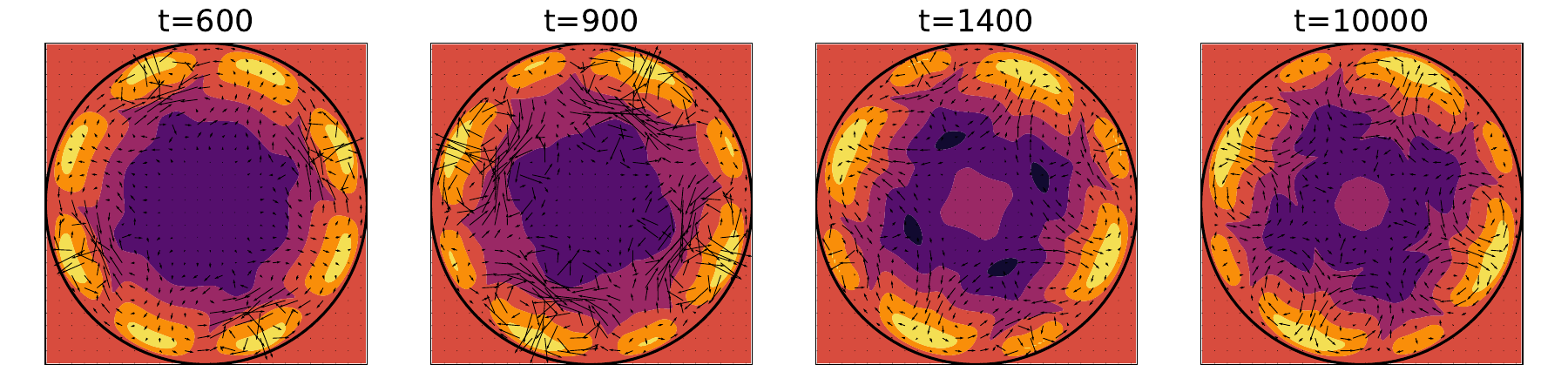}\\
\centerline{(d)}
\includegraphics[width=\linewidth,clip]{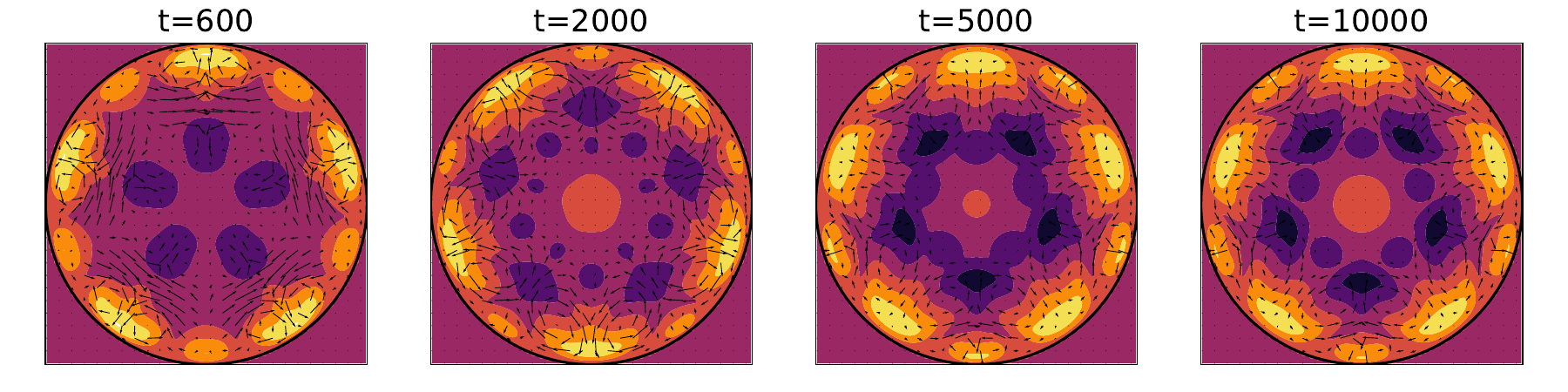}\\
%
\centerline{(e)}
\includegraphics[width=\linewidth,clip]{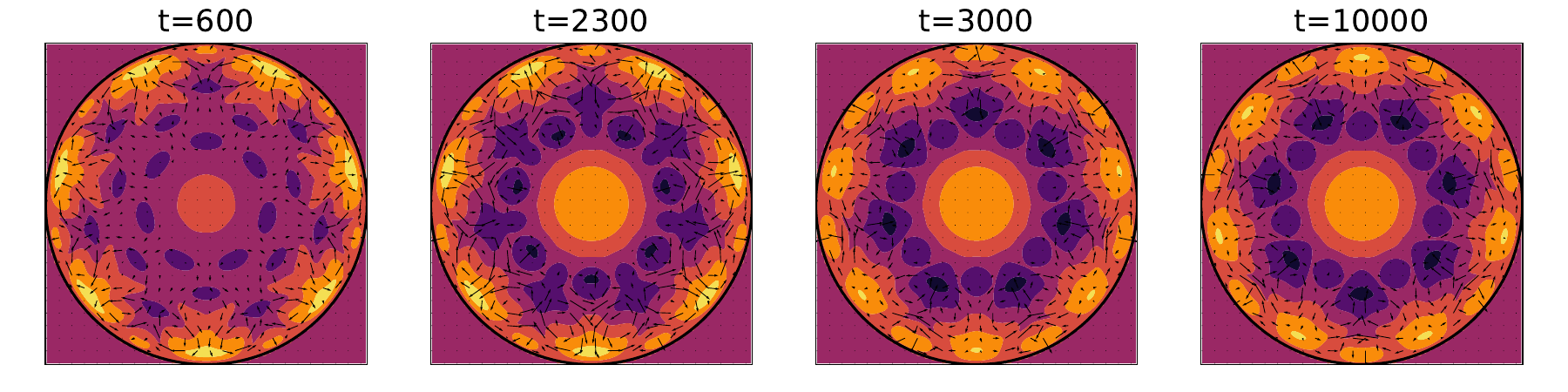}\\
\caption{
Snapshots of the $(r,\theta)$ plane at $z=0$ in the frame of reference translating with $c_z$ of the travelling waves for the MTDF cases outlined in table \ref{tab:2}. 
From top to bottom (a) \textsf{S2U} (b) \textsf{ML} (c) \textsf{N4U} (d) \textsf{N5} and (e) \textsf{N7}. 
Coloured contours represent $u_z$ (each with 7 contours evenly spaced in $[-0.45,0.45]$ except \textsf{S2U} which uses $[-0.35,0.35]$) and arrows to represent the in-plane velocity vector. 
Final snapshot at $t=10,000$ shows the stabilised state.
Supplementary materials contain the movies for these examples.
}
\label{fig:MTDF_xy_all}
\end{figure}
\subsection{Travelling waves \textsf{S2U, ML, N3, N4U, N5} and \textsf{N7}}\label{sec:TWs}

Having shown that \textsf{UB} is able to be stabilised from the turbulent state by our adaptive MTDF approach, we now demonstrate the generality of these results by presenting the stabilisation of the other travelling waves outlined in table \ref{tab:1}; 
the stabilising parameters can be found in table \ref{tab:2}. 
\begin{table}
\begin{center}
\begin{tabular}{lcccccccccccc}
sol.  & sym.& $Re$ & $\alpha$ & $\tau_1$ & $\tau_2$ & $\tau_3$ & $\tau_4$ & $G_1^{\max}$ & $G_{i\neq 1}^{\max}$ & $\gamma^G$  & $c_z(0)$ & $\gamma_s$ \\
\textsf{UB}    &  $S$,$Z_2$ & $3000$ & 1.25     &  2 & 9 & 17 & 33 & 0.1 & 0 & 0.1 & 0.65 & 0.1\\
\textsf{ML}    &  $S$,$Z_2$ & $3000$ & 1.25     &  2 & 9 & 33 & 150 & 0.1 & 0.01 & 0.5 & 0.71& 0.1\\
\textsf{S2U}    &  $S$ & $2400$ & 1.25     &  2 & 9 & 17 & 33 & 0.1 & 0.01 & 0.1 & 0.65& 0.1\\
\textsf{N4U}    &  $S$,$Z_4$ & $2500$ & 1.7 &  2 & 9 & 17 & 33 & 0.1 & 0.01 & 0.1 & 0.5 & 0.1\\
\textsf{N5}    &  $S$,$Z_5$ & $2500$ & 2 &  2 & 9 & 17 & 33 & 0.1 & 0.01 & 0.1 & 0.47& 0.01\\
\textsf{N7}    &  $S$,$Z_7$ & $3500$ & 3 &  2 & 9 & 17 & 33 & 0.1 & 0.01 & 0.1 & 0.4& 0.01
\end{tabular}
\caption{
Table summarising the parameters used in the four-term MTDF stabilisation of the travelling waves discussed. In all cases the start time $t_s=1000$ is used, $a=0.1,\,b=100$ for the sigmoid initialisation of $G_i$ with adaptivity started at $t=2000$ for all terms in all cases. Adaptivity of the translation (or phase speed) is initiated at $t=5000$, with initial value shown in the table as $c_z(0)$.
}
\label{tab:2}
\end{center}
\end{table}
First we tackle \textsf{S2U} at $Re=2400$ in the $S$ symmetry subspace where the solution has one pair of unstable eigenvalues $\lambda_{\pm}=0.14\pm0.13\mathrm{i}$ (note in the full space this solution violates the odd-number condition, see table \ref{tab:1}). In theory \modDL{it should be possible to stabilise} this solution with a single TDF term, provided $\tau$ and $G$ take suitable values. However, as in the previous case, we continue with MTDF as though we did not know the stability information, and start from the turbulent attractor. We use four terms with the same choices for $\tau_i,$ i.e. 2, 9, 17 and 33, as before. In the course of calibrating our adaptive methods it was noted that starting $G_i$ from 0 could lead to some slight instability; individual ``speed-gradients'' begin with a large value (not only when targeting this solution but in general). A simple way to avoid this is to begin with small non-zero gains before starting to adapt, but again initialised with a sigmoid function. 
In this case we keep all other parameters the same as the four-term \textsf{UB} case of figure \ref{fig:MTDF_turb_Re3000}, only with $G_i^{\max}=0.01$ for $i=2,3,4,$ and in the $S$ subspace, see table \ref{tab:2}. \modDL{The stabilisation of \textsf{S2U} appears quite similar to that of \textsf{UB}. 
Figure \ref{fig:MTDF_turb_all} (a) shows rapid stabilisation with the residual becoming very small $O(10^{-11}).$ Figure \ref{fig:MTDF_turb_all} (b) shows normalised energy input rate which demonstrates the large amplitude turbulent fluctuations before MTDF is activated and settling onto the final constant value at late times. It is observed that the residual only falls to small values once $c_z$ is dynamically adjusted to its exact value, shown in figure \ref{fig:MTDF_turb_all} (c). 
In this example the gains, plotted in figure \ref{fig:MTDF_turb_all} (d), do not undergo any significant dynamical adjustment. } Figure \ref{fig:MTDF_xy_all}(a) shows snapshots of this stabilisation \modDL{in the $(r,\theta)$ plane}.

\textsf{ML} is quite weakly unstable, in the $(S,Z_2)$ subspace at $Re=3000$ and 
the solution has only one unstable direction with $\lambda_{\pm}=0.0087\pm 0.019\mathrm{i}.$ 
This provides a useful test case to demonstrate that in this subspace at this Reynolds number, multiple solutions can be stabilised by only adjusting the MTDF parameters without requiring special treatment of the initial condition. 
Note that the unstable eigenfrequency is very small for this solution which necessitates using larger delay periods, tests with similar parameters used to stabilise \textsf{UB} either restabilise \textsf{UB} or relaminarise. 
In this instance we choose $\tau_i\in{2,9,33,150}$ and initiate with $c_z(0) = 0.71,$ the only other difference from earlier cases is that $\gamma^G=0.5$ for all terms (see table \ref{tab:2}). 
As is shown in figure \ref{fig:MTDF_turb_all}, \textsf{ML}, despite being much less unstable than the other cases we have studied, shows weaker stabilisation. 
In addition we see that $G_4$ undergoes significant growth once the speed-gradient method is initiated at $t=2000,$ becoming the largest gain of the four. 
%
Only once $G_4$ has grown we observe the solution stabilising, indicating that this term is dominant in ensuring stabilisation. This fact is consistent with the frequency domain analysis. 
In retrospect a larger starting $G^{\max}_4$ is likely to improve the stabilisation of \textsf{ML}. Nevertheless the adaptive approach has been able to determine stabilising gains automatically. Figure \ref{fig:MTDF_xy_all}(b) shows snapshots of this stabilisation.
Noting that \textsf{ML} and \textsf{UB} have been successfully stabilised at the same Reynolds number, in the same symmetry subspace and with quite similar MTDF parameters, we have verified that taking the parameters used to stabilise \textsf{ML} (row 2 of table \ref{tab:2}) and changing only $c_z(0)=0.65$ results in the stabilisation of \textsf{UB}. In other words both solutions can be obtained varying only $c_z(0).$

In $(S,Z_3)$ \modDL{at $Re=2500$ and $\alpha=2.5$ 
 a travelling wave,} \textsf{N3},  is stable, meaning that stabilisation is not necessary. However we have confirmed that by applying the symmetry operator $SZ_3$ to the TDF terms, e.g. 
\begin{eqnarray}\label{eq:MTDF_sym}
\bm{F}_{\mathrm{MTDF}}(r,\theta,z,t) 
= \sum_i^N G_i(t) 
[SZ_3\mathcal{T}_{z}(s_z) 
\bm{u}(r,\theta,z,t-\tau_i)-\bm{u}(r,\theta,z,t)] \:,  
\end{eqnarray}
stabilisation can be obtained in the full-space with some arbitrary (small) $\tau$ and $G_i,$ from a turbulent initial history. As explained in \citet{lucas_2022_stabilization}, TDF or MTDF will simply drive the dynamics onto the symmetry subspace where the travelling wave is an attractor. This result indicates that the MTDF results shown earlier could be repeated in less-restrictive subspaces, with symmetry operators embedded in the MTDF terms, similarly to \citet{lucas_2022_stabilization}.

\textsf{N4U} is stabilised in the $(S,Z_4)$ subspace at $Re=2500,\,\alpha=1.7$ 
with the same MTDF parameters used in the \textsf{S2U} case only now with $c_z(0)=0.5,$ as outlined in table \ref{tab:2}. 
There are few significant differences in this case, 
$G_4$ and $G_1$ both grow to values $\approx 0.3$ with $G_3$ also becoming relatively large. As with earlier examples an analysis of the unstable eigenfrequencies would indicate which terms are necessary in this case, but in the interests of brevity we will not repeat a similar calculation here, noting that, even without this analysis, significant parameter tuning was not necessary to stabilise this solution. 
As with the other travelling waves, the residual $R_{\mathrm{tot}}$ only begins to fall to values for which MTDF may be considered non-invasive as the phase-speed converges to its final value, i.e. $t>5000$. Figure \ref{fig:MTDF_xy_all}(c) shows snapshots of this stabilisation.

\section{Conclusion}\label{sec:conclusion}
%
In this paper, we present the first successful non-invasive stabilisation of \modDL{highly unstable (having multiple unstable directions)} nonlinear travelling waves in a straight \modDL{cylindrical} pipe through 
the use of an improved control method involving time-delayed feedback (TDF). 
Our novel TDF protocol allows for the stabilisation 
of multiple nonlinear travelling waves, at a range of Reynolds numbers, in a variety of symmetry subspaces, from a generic turbulent history and with speculative control parameters. 
Furthermore, our development of the TDF method has led to 
a deeper understanding of the principles governing time-delayed feedback in terms of an ``approximate'' linear stability analysis and frequency-domain analysis (see e.g., \S~\ref{SS:STDF_validation}). 
We have shown that the effect of TDF and MTDF on the unstable part of solution's eigenvalue spectrum can be approximated surprisingly well, enough to point a parameter study in the right direction. 
Moreover it has provided clearer insight into the frequency domain interpretation of the control method, which in turn gives a helpful means to choose delay periods in these cases. 
Finally we have shown that, if stabilisation is nearly achieved, it can be possible to diagnose eigenfrequencies and hence pick more appropriate time-delays, without the need for an {\textit{a priori}} stability analysis of the target solution. 
%
%

In order to enhance the performance of our TDF method, we implemented several optimisation methods that \modTY{enable} the feedback term(s) to vanish, as depicted in figure~\ref{fig:MTDF_turb_Re3000}b. 
Because the bulk flow in pipe flow is driven by an imposed axial pressure gradient, 
all invariant solutions take the form of relative solutions, such as travelling waves and relative periodic orbits. 
Therefore, applying a translation operator to the delay term(s) is essential. 
To achieve this, we utilised the adaptive shift method (see \S~\ref{SS:STDF}),
which dynamically adjusts the translations to match the phase speeds 
of the target solution through a simple ODE (\ref{eq:ode_shift}). 
We have demonstrated here, for the first time, that by changing the initial condition for $s_z,$ or $c_z,$ different travelling waves can be stabilised.

In the chaotic pipe flow system, 
we find that multiple time-delayed feedback control (MTDF) 
is effective at improving the control's ability to stabilise a wide range of unstable eigenfrequencies, as shown in \S~\ref{sec:MTDF}. 
A helpful consequence of MTDF is that it will serve to damp very slow temporal oscillations, which are typical in pipe flow \citep{shih_2016_ecological}. 
As demonstrated in the results of \S~\ref{SS:TDFresults} and \S~\ref{sec:TWs} 
successful stabilisations are found when MTDF is initiated with a short time delay ($\tau_1=2$ in all our cases), which acts for some time period before the rest of the terms become active. 
The effect is to suppress slow oscillations, importantly without relaminarising, 
giving the remaining terms a better chance at controlling the target travelling wave. 
In other words this initial delay is an effective means to widen the basin of attraction. 
In some cases $\tau_1$ contributes directly to altering the linear stability of the travelling waves, in others it only provides a nonlinear effect.
%
%

We have also sought to avoid expensive parametric studies. 
To achieve this, 
we have introduced the adaptive gain method (\S~\ref{SS:adaptG}) into our TDF protocol. 
This method automates our search for an appropriate gain, 
thereby avoiding an exhaustive parametric search for $G_i$. 
We have observed this approach to be highly effective, for instance in \S~\ref{sec:TWs} 
when stabilising \textsf{ML} the longest time delay gain $G_4$ is observed to grow significantly, compared to the other terms, showing that $\tau=150$ was important in ensuring stabilisation. 
This is consistent with our frequency analysis when noticing that 
this travelling wave has a very large unstable eigenperiod of around 330. 
%

During this work we have demonstrated that TDF, or more accurately MTDF, can stabilise multiple states at the same parameter values. In other words multiple attractors can coexist; \textsf{UB} and \textsf{ML} are stabilised varying only the initial $c_z.$ In this example the two states are relatively well separated in phase-space (upper and lower branch solutions, see figure \ref{fig:bif}) so it is perhaps surprising that \textsf{ML} is stabilised from a turbulent initial condition. This is a promising result as it demonstrates that the method does not require significant intervention to move from one solution to the next, however it does opens up a number of interesting questions. In particular, how might one design a systematic search, or data-driven approach, to explore basins of attraction of various potential solutions? For instance \textsf{UB} at $Re=2400$ in the $S$ subspace is highly unstable and has 9 complex pairs of eigenvalues, in theory this could be stabilised by MTDF, however it will ``compete'' with \textsf{S2U} which is much less unstable (1 unstable direction), is also upper branch and has a very similar phase speed (see table \ref{tab:1}). If any of the 9 eigenvalues has a particularly small eigenfrequency, necessitating a long time-delay, then close proximity to the solution is likely to be necessary for any stabilisation to be successful. 
We have also seen that trialling various subspaces and/or embedding symmetry operators into TDF terms is a useful way to obtain different solutions, avoid the odd-number issue and avoid dealing with multiple attractors. However we have not conducted an exhaustive search of all possible subspaces, pipe lengths and $Re.$ \modDL{We expect that travelling wave solutions, or even relative periodic orbits, can be stabilised with TDF, at a wide range of Reynolds numbers, pipe lengths, subspaces not to mention in other wall-bounded shear flows or systems with additional physics such as stratification or rotation where new dynamically important temporal frequencies arise.}

\modDL{The experimental applicability of these results is not immediately obvious. In order to address this, at least in a proof-of-concept manner, we have trialed a handful of cases where the control force $\bm{F}_{\textrm{TDF}}$ is premultiplied by the Kronecker delta function $\delta_{r_i=r_{N-1}}$. In other words TDF is no longer acting as a full-state control method but is applied only to the cylinder of grid points adjacent to the wall. For instance the stabilisation of \textsf{UB} shown in figure \ref{fig:Q} ($\tau=5,\, G=0.1$) is successful with this spatially localised control without any further parameter tuning (figures not shown for brevity). The fact that TDF can be successful here, when only applied to a subset of the degrees of freedom in the problem, gives some optimism that an experimental version of TDF may be successful.  It is also worth noting that TDF has been successful in an experimental Taylor-Couette flow \citep{Luthje:2001do} by applying a boundary forcing.}

In this paper, we have not tackled the odd-number 
limitation \citep{Nakajima1997}, which is a contentious issue in the TDF literature. 
Further improvements of this method are necessary 
in order to stabilise odd-number solutions. 
One possibility is the half-period TDF \citep{Nakajima_1998} which is similar in spirit to the use of symmetries here and in \cite{lucas_2022_stabilization}. 
Another is ``act-and-wait'' TDF \citep{pyragas_2018_act-and-wait,pyragas_2019_state-dependent}, where a time-dependent switching of the TDF gain is applied meaning uncontrolled dynamics are always used in the delay period, or ``unstable ETDF'' \citep{Pyragas:2001ch}, where an additional unstable degree of freedom is introduced into the problem to create an even pair of exponents. We have demonstrated that there is some potential for this method in controlling nonlinear states in spatiotemporal chaos, which will hopefully serve as motivation for further developments tackling both the odd-number issue and even higher dimensional problems at large Reynolds number and in large domains.
Potentially the most promising avenue for TDF in fluid flows is as a control method in a real physical system where a non-trivial flow state, perhaps of a specific dissipation or mixing rate, is desired but full spatiotemporal chaos is not. We have shown that with careful application of a time-delayed approach various kinds of target solution can be relatively easily obtained with minimal intervention. 

%
\backsection[Funding]{This work is supported by EPSRC New Investigator Award EP/S037055, “Stabilisation of exact coherent structures in fluid turbulence.”}

\backsection[Declaration of interests]{The authors report no conflict of interest.}
\bibliography{papers}
\bibliographystyle{jfm}
\end{document}